\documentstyle[epsfig,aps]{revtex} 
\begin{document}
\tighten
\draft

\def\D{\Delta}
\def\Da{\Delta_{(a)}}
\def\Dn{\Delta_\nu}
\def\Dp{\Delta_\gamma}
\def\Db{\Delta_b}
\def\Dc{\Delta_c}

\def\W{{V}}
\def\Vn{V_\nu}
\def\Vc{V_c}
\def\Vbg{V_{b\gamma}}

\def\Dno{\Delta^{(0)}_\nu}
\def\Dpo{\Delta^{(0)}_\gamma}
\def\Dbo{\Delta^{(0)}_b}
\def\Dco{\Delta^{(0)}_c}
\def\Psio{{\Psi^{(0)}}}
\def\Phio{{\Phi^{(0)}}}
\def\Vno{\Vn^{(1)}}
\def\Vco{\Vc^{(1)}}
\def\Vbgo{\Vbg^{(1)}}

\def\k{{\bf k}}
\def\x{{\bf x}}

\def\dc{{\delta}}
\def\h{{\cal H}}

\def\Omb{\Omega_b}
\def\Omc{\Omega_c}
\def\Omp{\Omega_\gamma}
\def\Omn{\Omega_\nu}

\def\md{^{{\rm MD}}}
\def\rd{^{{\rm RD}}}

\newcommand{\ddd}{{\mathrm d}}
\newcommand{\Hconf}{{\mathcal H}}

\newcommand{\FIG}[1]{Fig.~{#1}}
\newcommand{\FFIG}[1]{Fig.~{#1}}
\newcommand{\FIGS}[1]{Figs.~{#1}}
\newcommand{\FFIGS}[1]{Figs.~{#1}}
\newcommand{\EQ}[1]{eq.~(#1)}
\newcommand{\EQNS}[1]{eqns.~(#1)}

\newcommand{\IE}{i.e.}
\newcommand{\EG}{e.g.}

\title{Correlated mixtures of adiabatic and isocurvature cosmological 
 perturbations}
\author{David Langlois, Alain Riazuelo}
\address{D\'epartement d'Astrophysique Relativiste et de Cosmologie,\\
Centre National de la Recherche Scientifique,\\
Observatoire de Paris, 92195 Meudon Cedex, France}
\date{\today}
\maketitle

\begin{abstract} 

We examine the consequences of the existence of correlated mixtures of
adiabatic and isocurvature perturbations on the CMB and large scale
structure. In particular, we consider the four types of ``elementary''
totally correlated hybrid initial conditions, where only one of the
four matter species (photons, baryons, neutrinos, CDM) deviates from
adiabaticity.  We then study the height and position of the acoustic
peaks with respect to the large angular scale plateau as a function of
the isocurvature to adiabatic ratio.
\end{abstract}

\def\beq{\begin{equation}}
\def\eeq{\end{equation}}
\def\d{\delta}

\section{Introduction}

In the near future, a lot of data about the anisotropies
of the Cosmic Microwave Background (CMB) will be available to
cosmologists, notably thanks to balloon experiments and the planned
satellites MAP~\cite{map} and Planck~\cite{planck}.  What will be
remarkable is the expected high resolution and sensitivity of these
experiments, which may turn cosmology into a high precision activity.

One of the hopes of cosmologists is to be able to determine from these
data the cosmological parameters describing the geometry and matter
contents of our Universe. In this respect, it is important to stress
that the fluctuations that are and that will be measured, result,
according to our current understanding, from a {\it combination of the
primordial perturbations and of the cosmological parameters}.  In the
preparation of the future data analysis, one should be careful to
avoid oversimplification {\it a priori} of the primordial
perturbations and not to stick to the simplest one-scalar field
inflation model. After all, the early universe is the period in the
history of the Universe where the physics is the least known.

A more general description of the primordial perturbations may
therefore be needed to be able to interpret the future data.  In this
perspective, the aim of this work is to examine the consequences of
the existence of isocurvature perturbations in addition to the usual
adiabatic perturbations. Such studies have already been performed in
the case of {\it independent} mixtures of adiabatic and isocurvature
perturbations~\cite{sbg96}-\cite{pgb99}. This is why we will focus our
attention on {\it correlated} mixtures of adiabatic and isocurvature
perturbations.  The possibility of such primordial perturbations is
motivated by the recent work of one of us, which showed that the
simplest model of multiple inflation, a model with two massive
non-interacting scalar fields, can produce such correlated
mixtures~\cite{l99}.

Isocurvature perturbations are perturbations in the relative density
ratio between various species in the early universe, in contrast with
the more standard adiabatic (or isentropic) perturbations which are
perturbations in the total energy density with fixed particle number
ratios.  Primordial isocurvature perturbations are often ignored in
inflationary models. The main reason for this is that they are less
universal than adiabatic perturbations because, on one hand, they can
be produced only in multiple inflationary models~\cite{multinf}, and,
on the other hand, they do not necessarily survive until the present
epoch.

However, isocurvature perturbations have been shown to be of potential
importance in some specific models~: axions~\cite{ksy96},
\cite{burns}, \cite{kksy99}, \cite{eb86}, Affleck-Dine baryogenesis
mechanism~\cite{em98}, multiple field inflation~\cite{l99},
\cite{multinf}, \cite{lm97}, \cite{ps94}.

A priori, since the (not too old) Universe is filled with four
species, baryons, photons, neutrinos and dark matter (which will be
assumed to be cold here), several types of isocurvature perturbations
can be envisaged. For example, in the past, a model with isocurvature
baryon perturbations was proposed~\cite{peebles87}, although it does
not seem compatible with the data today~\cite{hbs95}. Most of the
recent models, however, contain cold dark matter (CDM) isocurvature
perturbations. A more general approach, including neutrino
isocurvature perturbations (and also isocurvature velocity
perturbations), was considered recently~\cite{bmt99}.  In the present
work, we will focus our attention, for simplicity, on primordial
perturbations where only one species deviates from adiabaticity, which
thus leaves room to four types of hybrid (\IE{} adiabatic plus
isocurvature) initial perturbations. These four ``elementary'' modes
will be systematically studied, without trying to make any connection
with specific early universe models.

As far as observational constraints are concerned, it has already been
established that a pure isocurvature scale-invariant spectrum must be
rejected because it predicts on large scales too large temperature
anisotropies with respect to density fluctuations~\cite{eb86}, but
other possibilities have been envisaged, like tilted isocurvature
perturbations.  The main trend, however, has been to study models with
a mixing of isocurvature and adiabatic perturbations. Confrontation of
these models with observational data, like CMB anisotropies and large
scale structures, seems to allow only for a small fraction of
isocurvature perturbations.  Future CMB measurements will also enable
us to put much tighter constraints on this kind of models.
 
It must be emphasized, however, that all these studies assumed {\it
independent} mixing of isocurvature and adiabatic perturbations.
While this assumption can be indeed justified in some specific early
universe models, it is certainly not an absolute rule, as it has been
shown in~\cite{l99}. It is thus the purpose of this paper to
investigate the consequences on observational quantities, namely the
large scale structure and the CMB anisotropies, of {\it correlated
mixtures of isocurvature and adiabatic} perturbations.  As it will be
shown, correlation gives more richness to hybrid perturbations.  For
example, while the first acoustic peak (relatively to the plateau) is
always lower for independent hybrid perturbations than for pure
adiabatic perturbations, it can be either lower or higher with
correlated hybrid perturbations.

The plan of the paper will be the following. In the next section
(\S\ref{sec_adi_iso}), we recall the basic definitions of isocurvature
and adiabatic perturbations and introduce hybrid perturbations. Then,
\S\ref{sec_corr} will discuss the notion of correlation between
isocurvature and adiabatic perturbations.  In \S\ref{sec_long_wave},
we will begin the systematic analysis of the correlated hybrid
perturbations by considering the long wavelength modes, which can be
solved analytically. \S\ref{sec_numeric} is devoted to the numerical
investigation for all cosmological scales. Finally, we present our
conclusions in \S\ref{sec_conc}. There is also an appendix, which
details the evolution equations and gives the full solution for the
long wavelength modes.

\section{Adiabatic and isocurvature  perturbations}
\label{sec_adi_iso}

In this section, we will define precisely the notion of adiabatic and
isocurvature perturbations, and introduce the notation that will be
used throughout this paper. Only perturbations of the scalar type will
be considered here.

The perturbations of the spacetime geometry will be described by two
scalar potentials, $\Phi$ and $\Psi$, which appear in the linear
perturbation of the (flat) Friedmann-Lema\^\i{}tre-Robertson-Walker
(FLRW) metric,
\begin{equation}
ds^2=-a^2(\eta)(1+2\Phi)\ddd \eta^2+
      a^2(\eta)(1-2\Psi)\delta_{ij}\ddd x^i \ddd x^j, 
\label{metric}
\end{equation}
choice which corresponds to the longitudinal gauge ($\eta$ is the
conformal time and $\delta_{ij}$ is the Kronecker symbol).

For matter, we will consider four different species~: two relativistic
species, photons and massless neutrinos; two non-relativistic species,
baryons and cold dark matter (CDM). Their respective energy density
contrasts will be denoted $\d_\gamma$, $\d_\nu$, $\d_b$ and $\d_c$
($\d_A\equiv \d\rho_A/\rho_A$).

Before defining adiabatic and isocurvature perturbations, let us
introduce the idea of entropy perturbation, which can be defined for
any pair of components, $A$ and $B$, by the expression
\beq
S_{A,B}\equiv {\dc n_A\over n_A}-{\dc n_B\over n_B},
\eeq
where $n_A$ represents the particle number density for the species
$A$. When the equation of state for a given species is such that
$w\equiv p/\rho= {\rm Const}$, then one can reexpress the entropy
perturbation in terms of the density contrast, in the form
\beq
S_{A,B}\equiv {\d_A\over 1+w_A}-{\d_B\over 1+w_B}.
\eeq

By definition, an adiabatic (or isentropic) perturbation corresponds
to the case where all the entropy perturbations are zero.  For our
four components, this can be expressed as~:
\beq
\d_c=\d_b=\frac{3}{4}\d_\nu=\frac{3}{4}\d_\gamma\equiv \d_{(a)},
\label{adiabatic}
\eeq
where $\d_{(a)}$ will denote the common value (up to the adiabatic index 
of the equation of state) of the density contrasts.
An adiabatic perturbation is thus characterized by a unique amplitude, 
which can be $\d_{(a)}$ but  which is usually given, for convenience, in 
terms of the gravitational potential $\Psi$, which can be directly 
relation to $\d_{(a)}$ through Poisson's equation (see appendix).

An isocurvature perturbation, as indicated by its name, corresponds to
a perturbation for which the gravitational potential perturbation is
zero (or approximately zero). To get a non-trivial isocurvature
perturbation, one must therefore have several components and at least
one non-vanishing entropy perturbation $S_{A,B}$.  For $N$ species,
there will be one adiabatic mode and $N-1$ independent isocurvature
modes (there is also the possibility to have velocity-type
isocurvature modes, see~\cite{bmt99}).  It is to be noticed that the
adiabatic or isocurvature character of perturbations is not
time-invariant. In the cosmological context, when one talks about
adiabatic or isocurvature perturbations, one implicitly assumes that
this property corresponds to the {\it initial} state of the
perturbations, which means deep in the radiation era when the
wavelength of the perturbation is much larger than the Hubble radius.

In the following, we shall consider primordial perturbations which are
hybrid, \IE{} which are a sum of adiabatic and isocurvature modes.  In
order to simplify the exploration of the parameter space, we will
restrict our attention to the case of perturbations for which all
entropy perturbations within three of the species are zero whereas the
last species, $X$ say, departs from adiabaticity. This means that
three of the four species will satisfy the above relation
(\ref{adiabatic}) while the contrast density for the remaining species
$X$ will be written in the form
\beq
{\d_X\over 1+w_X}=S_X+\d_{(a)},
\eeq
expression which defines the isocurvature perturbation $S_X$
associated with the species $X$. Varying $X$, one can construct four
hybrid perturbations of this type.

\section{Correlations}
\label{sec_corr}

In cosmology, perturbations are treated as homogeneous and isotropic
random fields. It is convenient to deal with them in Fourier space
rather than ordinary space, and all quantities defined previously can
be transformed into their Fourier components, according to the
relation (since we work only in flat space)~:
\begin{equation}
f_{\k}=\int{\ddd^3\x\over (2\pi)^{3/2}} e^{-i\k.\x} f(\x).
\end{equation}

Primordial perturbations are usually assumed to be Gaussian, in which
case their statistical properties can be summarized simply in terms of
their power spectrum, defined for a quantity $f$ by
\begin{equation}
\langle f_{\k}f_{\k'}^*\rangle = 2\pi^2 k^{-3}{\cal P}_f(k)
\delta(\k-\k').\label{spectrum}
\end{equation}
When primordial perturbations are described by {\it several
quantities}, such as would be the case if one has a mixture of
adiabatic and isocurvature perturbations, one can also define, for any
pair of random fields $f$ and $g$, a {\it covariance spectrum} ${\cal
C}_{f,g}(k)$ by the following expression~:
\begin{equation}
\Re\langle f_\k g^*_{\k'}\rangle  = 2\pi^2 k^{-3} {\cal C}_{f,g}(k)
\delta(\k-\k').
\end{equation}
Correlation between $f$ and $g$ can also be expressed in terms of a
{\it correlation spectrum} ${\tilde{\cal C}}_{f,g}(k)$ obtained by
normalizing ${\cal C}_{f,g}(k)$~:
\beq
{\tilde{\cal C}}_{f,g}(k)={{\cal C}_{f,g}(k)\over \sqrt{{\cal P}_f(k)}
\sqrt{{\cal P}_g(k)}}. \label{correlation}
\eeq

In the present work, we are especially interested in the possible
correlation between adiabatic and isocurvature primordial
perturbations, \IE{} ${\tilde{\cal C}}_{\Phi,S}(k)$.  Until very
recently, only independent mixtures, \IE{} with vanishing correlation,
were considered in the literature.  This statistical independence
means that the quantities $\Phi$ and $S$ can be expressed as
\beq
\Phi={\cal P}_\Phi^{1/2} e_1, \quad S={\cal P}_S^{1/2} e_2,
\label{independence}
\eeq
where $e_1$ and $e_2$ are {\it independent} normalized centered
Gaussian random fields (\IE{} such that $\langle e_i(\k)\rangle=0$,
$\langle e_i(\k)e_j^*(\k')\rangle=\delta_{ij} \delta(\k-\k')$, for
$i,j=1,2$), and where the subscripts $\k$ are implicit, as it will be
the case in the rest of this paper. With the assumption
(\ref{independence}), one obtains immediately vanishing covariance and
correlation spectra.

However, as it was shown in a specific model of double
inflation~\cite{l99}, one can also envisage models of the early
universe where {\it correlated} primordial perturbations are
generated.  To be more specific, this would be the case if one
imagines several independent stochastic processes taking place in the
early Universe, which contribute {\it both } to adiabatic and
isocurvature perturbations, \IE such that
\beq
\Phi=\sum_i \Phi_i e_i, \quad S=\sum_i S_i e_i, 
\eeq
where the $e_i$ are {\it independent} normalized centered Gaussian
random fields. 
In the specific example of~\cite{l99}, there were two
independent random fields, generated by the quantum fluctuations of
two scalar fields.

In the present article, our goal will be to study systematically the
consequences of a {\it totally correlated mixture} of adiabatic and
isocurvature perturbations, \IE{} primordial perturbations which can
be written in terms of one {\it single} random field.  Of course, the
consequences of more general initial conditions can then be obtained
by simply adding the spectra (to get the total density power spectrum
or the total temperature fluctuation multipole spectrum) of several
totally correlated initial conditions.

\section{Long wavelength analysis}
\label{sec_long_wave}

As shown in the appendix, it is possible to solve analytically for the
long wavelength perturbations. Totally correlated perturbations can be
defined by two primordial quantities~: the gravitational potential
perturbation deep in the radiation era, denoted $\hat\Phi$, and the
entropy perturbation denoted $S_X$ (the index $X$ depends on the
species which departs from adiabaticity as explained in Section
\ref{sec_adi_iso}). It is then possible, using the expressions of the
appendix, to compute observational quantities at the time of last
scattering as functions of the two primordial quantities $\hat\Phi$
and $S_X$, {\it for scales larger than the Hubble radius at the time
of last scattering}.

In this section, we will use, instead of the energy density contrasts
$\d_A$ defined previously in the longitudinal gauge (\ref{metric}),
the slightly redefined energy density contrasts (defined in the
flat-slicing gauge)~:
\beq
\Dp=\d_\gamma-4\Psi, \quad \Dn=\d_\nu-4\Psi, \quad \Db=\d_b-3\Psi,
\quad \Dc=\d_c-3\Psi.
\eeq
The reason to use these quantities is essentially that the
conservation equations look much simpler (see appendix). Moreover, it
is to be noticed that, with these new density contrasts, the
definitions of adiabatic and isocurvature fluctuations keep exactly
the same form.  In other words, a purely adiabatic perturbation, as
defined by (\ref{adiabatic}), will also be characterized by~:
\beq
\Db=\Dc=\frac{3}{4}\Dn=\frac{3}{4}\Dp\equiv\Da.
\eeq
For a mixed perturbation, with the species $X$ deviating from
adiabaticity, the same relation will hold for the three species other
than $X$, and the density of the latter will be given by
\beq
{\D_X\over 1+w_X}=S_X+\Da.
\eeq

Our purpose will now be to express the observable quantities for long
wavelength modes, namely the gravitational potential in the matter era
and the temperature anisotropies, in terms of the primordial
quantities $\hat \Phi$ and $S_X$.  This will be possible by using the
two following relations, which are demonstrated in the appendix. The
first relation gives the metric perturbation in terms of the
primordial density contrast, during the radiation era,
\beq
\Phi_{{\rm rad}}\equiv 
\hat \Phi=-\frac{1}{4}\left(3+\frac{4}{5}\Omn\rd\right)^{-1}\left[
2\left(1-\frac{4}{5}\Omn\rd\right)\Omp\rd\Dp
+\frac{2}{5}\left(9-4\Omn\rd\right)
\Omn\rd\Dn\right], \label{phi_rad}
\eeq
the second one being the analogous equation during the matter era,
\beq
\Phi_{{\rm matter}}=-\frac{1}{5}\left(\Omb\md\Db+\Omc\md\Dc \right). 
\label{phi_mat}
\eeq
Note that, in the above equations, $\Omn\rd$ and $\Omp\rd$ are taken
in the radiation era, whereas $\Omb\md$ and $\Omc\md$ correspond to
their values in the matter era.

Our hybrid perturbations can be specified either by the pair $(\hat
\Phi, S_X)$ or the pair $( \Da, S_X)$, the relation between the two
following immediately from the relation (\ref{phi_rad}).  For a purely
adiabatic perturbation, it is easy to see, using (\ref{phi_rad}), that
\beq
\Da=-\frac{3}{2}\left(3+\frac{4}{5}\Omn\rd\right)\hat 
\Phi\equiv \alpha \hat\Phi.
\label{Da}
\eeq
For a mixed perturbation, there will be in general an additional term
proportional to $S_X$.  In the case of the baryons and CDM, the
expression for $\Da$ is the same as the adiabatic case, simply because
$\Db$ and $\Dc$ do not appear in (\ref{phi_rad}).  For the
relativistic species, one obtains
\beq
\Da=-\frac{3}{2}\left(3+\frac{4}{5}\Omn\rd\right)\hat \Phi 
-\left(1-\frac{4}{5}\Omn\rd\right)\Omp\rd S_\gamma 
\equiv \alpha \hat\Phi+\beta_\gamma  S_\gamma
\label{Da_p}
\eeq
for a photon-type hybrid perturbation and
\beq
\Da=-\frac{3}{2}\left(3+\frac{4}{5}\Omn\rd\right)\hat \Phi 
-\frac{1}{5}\left(9-4\Omn\rd\right)\Omn\rd S_\nu
 \equiv \alpha \hat\Phi+\beta_\nu  S_\nu
\label{Da_n}
\eeq
for a neutrino-type hybrid perturbation.

Substituting in (\ref{phi_mat}) the expressions of $\Db$ and $\Dc$ in
terms of $\Da$ and $S_X$, thus in terms of $\hat \Phi$ and $S_X$, it
is then possible to find the gravitational potential perturbation
during the matter era.  For a purely adiabatic perturbation, one finds
\beq
\Phi_{{\rm adiab}}=\frac{3}{10}\left(3+\frac{4}{5}\Omn\rd\right)\hat \Phi.
\label{Phi_adiab}
\eeq
One recognizes the standard transfer coefficient of $9/10$ if one
ignores the anisotropic pressure of neutrinos (see
\EG~\cite{ll}). Here, we have its generalization, which is
numerically very close to $1$, when the anisotropic pressure is taken
into account.  For a hybrid perturbation, the gravitational potential
perturbation during the matter era, will be of the form
\beq
\Phi=\Phi_{{\rm adiab}}+\Phi_{{\rm isoc}}, \label{Phi_decompos}
\eeq
where $\Phi_{{\rm adiab}}$ corresponds to the term proportional to
$\hat\Phi$, which is, in all cases, given by the same expression
(\ref{Phi_adiab}), and $\Phi_{{\rm isoc}}$ is the term proportional to
$S_X$ , whose explicit expression depends on the particular species
considered.  For baryons and CDM, its form is simply
\beq
\Phi_{{\rm isoc}}=-
\frac{1}{5}\Omega_X\md S_X, \quad X=b,c.
\eeq
For a photon-type mixed perturbation, one finds 
\beq
\Phi_{{\rm isoc}}=
\frac{1}{5}\left(1-\frac{4}{5}\Omn\rd\right)\Omp\rd S_\gamma,
\eeq
and finally, for a neutrino-type hybrid perturbation, one gets
\beq
\Phi_{{\rm isoc}}=\frac{1}{25}\left(9-4\Omn\rd\right)\Omn\rd S_\nu.
\eeq
The decomposition (\ref{Phi_decompos}) expresses the fact that a
primordial isocurvature perturbation will also contribute, {\it in the
matter era}, to the potential perturbation, whereas it is of course
not the case in the radiation era. This illustrates, once more, that
the separation between adiabatic and isocurvature modes is not
conserved during time evolution.

Let us now evalute the contribution of the primordial perturbations to
the CMB temperature anisotropies, here only for large angular scales
since we are restricted to long-wavelength perturbations. Neglecting a
local monopole and dipole contribution, the temperature anisotropies,
due to scalar perturbations, are approximatively given by~:
(see~\cite{sw67}, \cite{panek86})
\beq
\frac{\Delta T}{T}=
\frac{1}{4}{\Dp}_{{\rm LSS}}+(\Phi+\Psi)_{{\rm LSS}}-
e^i\partial_i(V_{{\rm LSS}}) +\int_{\eta_{{\rm LSS}}}^{\eta_0}
\left(\dot\Phi+\dot\Psi\right) \ddd \lambda,
\eeq
where $e^i$ is a spatial unit vector corresponding to the direction of
observation, the subscript LSS indicates that the quantities are
evaluated at the last scattering surface, a dot denotes a derivation
with respect to the conformal time $\eta$, $\eta_0$ is today's
conformal time, and the integral in the last term runs on the photon
line-of-sight. The contribution due to the first two terms is usually
called the Sachs-Wolfe term (SW), while the third term is called the
Doppler term and the last one the integrated Sachs-Wolfe term (ISW).
In general, but not always (see the pathological cases below), the SW
term is dominant for large angular scales. In terms of our variables,
the SW term can be written
\beq
\left(\frac{\Delta T}{T}\right)_{{\rm SW}}=
\frac{1}{4}\Dp+\Phi+\Psi\simeq \frac{1}{4}\Dp+2\Phi,
\eeq
where the quantities are evaluated at the last scattering surface,
assuming that last scattering occurs well in the matter era (in this
case $\Psi\simeq
\Phi$).
Using the expressions obtained above, it is now possible to express the 
SW term as a function of the primordial perturbations $\hat \Phi$ and 
$S_X$. As for the gravitational potential perturbation, one can decompose
this term into
\beq
\left(\frac{\Delta T}{T}\right)_{{\rm SW}}
=\left(\frac{\Delta T}{T}\right)_{{\rm adiab}} + \left(\frac{\Delta
T}{T}\right)_{{\rm isoc}},
\eeq
where the adiabatic component is the term proportional to $\hat \Phi$
and the isocurvature component is proportional to $S_X$. For all types
of hybrid perturbations, the adiabatic term is the same~:
\beq
\left(\frac{\Delta T}{T}\right)_{{\rm adiab}}=
\frac{1}{10}\left(3+\frac{4}{5}\Omn\rd\right)\hat \Phi.
\label{T_adiab}
\eeq
Note that one has ${\Delta T/ T}_{{\rm adiab}}=\Phi_{{\rm adiab}}/3$,
which reminiscent of the standard (adiabatic) Sachs-Wolfe term.  As
for the isocurvature term, it will depend on the particular type of
perturbation. For hybrid perturbations which are baryon or CDM
isocurvature, one finds
\beq 
\left(\frac{\Delta T}{T}\right)_{{\rm isoc}}=
-\frac{2}{5}\Omega_X\md S_X, \quad X=b, c.
\label{T_iso_c}
\eeq
Note that, for baryons and CDM, one has the relation $({\Delta T/
T})_{{\rm isoc}}=2\Phi_{{\rm isoc}}$, and by comparison with the
similar relation for the adiabtic terms, one recognizes the standard
statement in the literature that pure isocurvature perturbations (of
the baryon- or CDM-type) produce large scale temperature fluctuations
{\it six times bigger} than pure adiabatic perturbations.  For photon
isocurvature hybrid perturbation, one will get
\beq
\left(\frac{\Delta T}{T}\right)_{{\rm isoc}}=
\frac{1}{15}\left(6-\frac{9}{5}\Omn\rd+\frac{4}{5}\left(\Omn\rd\right)^2
\right) S_\gamma,
\eeq
whereas, for neutrino isocurvature hybrid perturbation, the expression
is
\beq
\left(\frac{\Delta T}{T}\right)_{{\rm isoc}}=
\frac{1}{75}\left(9-4\Omn\rd\right)\Omn S_\nu.
\eeq

\section{Numerical analysis}
\label{sec_numeric}

The present section will be devoted to the predictions of temperature
anisotropies, as well as large scale structure power spectrum, for
primordial correlated hybrid adiabatic and isocurvature perturbations.
We will keep fixed a certain number of parameters~: $\Omega_\Lambda =
0$, $\Omega_0 = 1$, $h_{100}=0.5$, $\Omega_b=0.05$, three species of
massless non-degenerate neutrinos (leading to $\Omp\rd =1/[1+(21/8)
(4/11)^{4/3}]\simeq 0.595$ and $\Omn\rd = 1-\Omp\rd\simeq 0.405$), and
standard recombination. The primordial perturbations will be assumed
to be scale-invariant.

\subsection{Temperature anisotropies}

\begin{figure}
\centering
% GNUPLOT: LaTeX picture with Postscript
\begingroup%
  \makeatletter%
  \newcommand{\GNUPLOTspecial}{%
    \@sanitize\catcode`\%=14\relax\special}%
  \setlength{\unitlength}{0.12bp}%
\begin{picture}(3600,2160)(0,0)%
\special{psfile=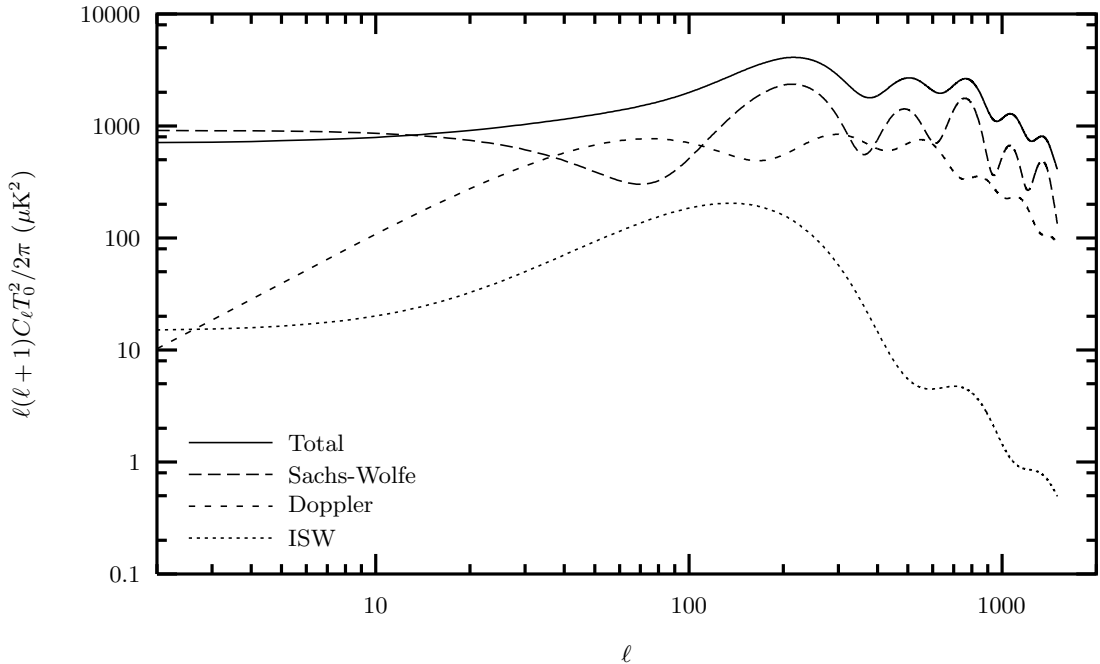 llx=0 lly=0 urx=600 ury=420 rwi=7200}
\put(913,413){\makebox(0,0)[l]{ISW}}%
\put(913,513){\makebox(0,0)[l]{Doppler}}%
\put(913,613){\makebox(0,0)[l]{Sachs-Wolfe}}%
\put(913,713){\makebox(0,0)[l]{Total}}%
\put(1975,50){\makebox(0,0){$\ell$}}%
\put(100,1180){%
\special{ps: gsave currentpoint currentpoint translate
270 rotate neg exch neg exch translate}%
\makebox(0,0)[b]{\shortstack{$\ell(\ell+1) C_\ell T_0^2 / 2 \pi$ ($\mu$K$^2$)}}%
\special{ps: currentpoint grestore moveto}%
}%
\put(3154,200){\makebox(0,0){1000}}%
\put(2171,200){\makebox(0,0){100}}%
\put(1187,200){\makebox(0,0){10}}%
\put(450,2060){\makebox(0,0)[r]{10000}}%
\put(450,1708){\makebox(0,0)[r]{1000}}%
\put(450,1356){\makebox(0,0)[r]{100}}%
\put(450,1004){\makebox(0,0)[r]{10}}%
\put(450,652){\makebox(0,0)[r]{1}}%
\put(450,300){\makebox(0,0)[r]{0.1}}%
\end{picture}%
\endgroup
 
\caption{CMB anisotropies in the pure adiabatic model ($\lambda =
0$). The solid line represents the total (scalar) contribution. The
Sachs-Wolfe, Doppler and Integrated Sachs-Wolfe contributions are
respectively represented in long-dashed, short-dashed, and dotted
lines. 
At large angular scales (low $\ell$), the total amplitude is essentially due 
to the Sachs-Wolfe contribution.}
\label{fig_adi}
\end{figure}

\begin{figure}
\centering
% GNUPLOT: LaTeX picture with Postscript
\begingroup%
  \makeatletter%
  \newcommand{\GNUPLOTspecial}{%
    \@sanitize\catcode`\%=14\relax\special}%
  \setlength{\unitlength}{0.12bp}%
\begin{picture}(3600,2160)(0,0)%
\special{psfile=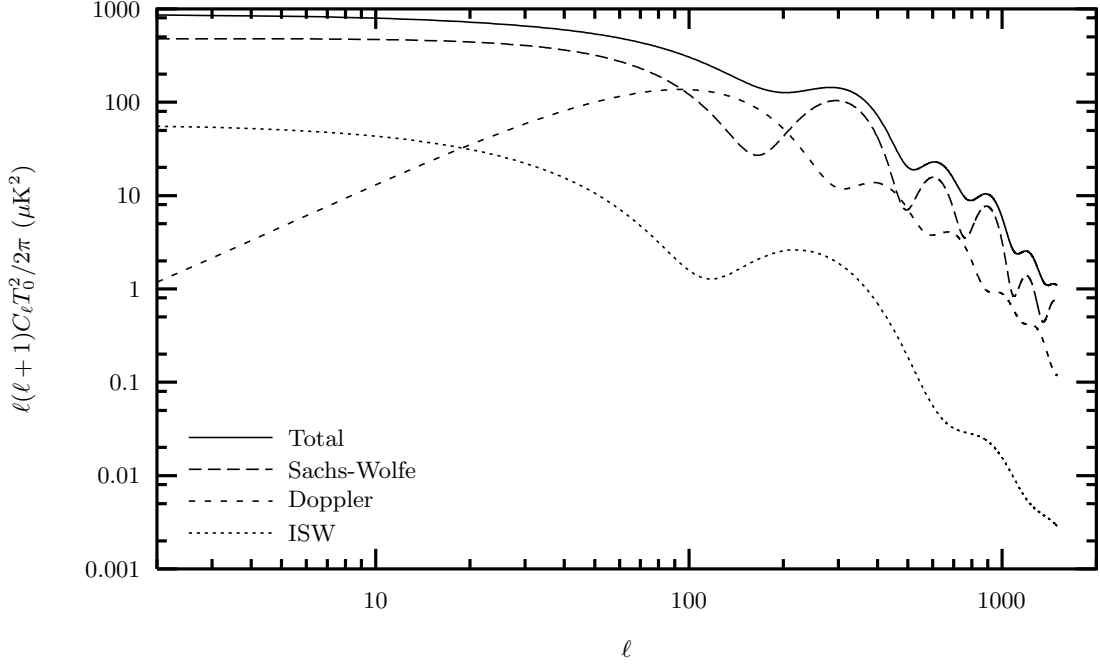 llx=0 lly=0 urx=600 ury=420 rwi=7200}
\put(913,413){\makebox(0,0)[l]{ISW}}%
\put(913,513){\makebox(0,0)[l]{Doppler}}%
\put(913,613){\makebox(0,0)[l]{Sachs-Wolfe}}%
\put(913,713){\makebox(0,0)[l]{Total}}%
\put(1975,50){\makebox(0,0){$\ell$}}%
\put(100,1180){%
\special{ps: gsave currentpoint currentpoint translate
270 rotate neg exch neg exch translate}%
\makebox(0,0)[b]{\shortstack{$\ell(\ell+1) C_\ell T_0^2 / 2 \pi$ ($\mu$K$^2$)}}%
\special{ps: currentpoint grestore moveto}%
}%
\put(3154,200){\makebox(0,0){1000}}%
\put(2171,200){\makebox(0,0){100}}%
\put(1187,200){\makebox(0,0){10}}%
\put(450,2060){\makebox(0,0)[r]{1000}}%
\put(450,1767){\makebox(0,0)[r]{100}}%
\put(450,1473){\makebox(0,0)[r]{10}}%
\put(450,1180){\makebox(0,0)[r]{1}}%
\put(450,887){\makebox(0,0)[r]{0.1}}%
\put(450,593){\makebox(0,0)[r]{0.01}}%
\put(450,300){\makebox(0,0)[r]{0.001}}%
\end{picture}%
\endgroup
 
\caption{CMB anisotropies in the pure isocurvature CDM model ($\lambda
= \pm\infty$). The solid line represent the total scalar
contribution. The Sachs-Wolfe (SW), Doppler and Integrated Sachs-Wolfe
contributions are respectively represented in long-dashed,
short-dashed, and dotted lines. Note  that the power at large scales 
(low $\ell$)  is higher than at the degree-scale.}
\label{fig_iso}
\end{figure}

As far as temperature anisotropies are concerned, isocurvature
perturbations can be distinguished from pure isocurvature
perturbations by a much larger plateau, as explained in the previous
section, with the consequence that the all the acoustic peaks will
appear smaller than this plateau. To show these two extreme behaviours
we have plotted on \FIG{\ref{fig_adi}} the case of pure adiabatic
initial conditions, and on \FIG{\ref{fig_iso}}, the case of pure
CDM-type isocurvature initial conditions.  We have also plotted, in
each case, the SW, Doppler and ISW contributions.  We have used, as
usual, the angular power spectrum for the temperature anisotropies,
defined by
\beq
C_\ell^{TT}=\langle |a_{\ell m}^T|^2\rangle,
\eeq
where the $a_{\ell m}^T$ are the multipole coefficients that appear in
the decomposition into spherical harmonics of the temperature
fluctuations, \IE
\beq
{\Delta T\over T}=\sum_{\ell,m}a_{\ell m}^T Y_{\ell m}.
\label{multipole}
\eeq

\begin{figure}
\centering
% GNUPLOT: LaTeX picture with Postscript
\begingroup%
  \makeatletter%
  \newcommand{\GNUPLOTspecial}{%
    \@sanitize\catcode`\%=14\relax\special}%
  \setlength{\unitlength}{0.12bp}%
\begin{picture}(3600,2160)(0,0)%
\special{psfile=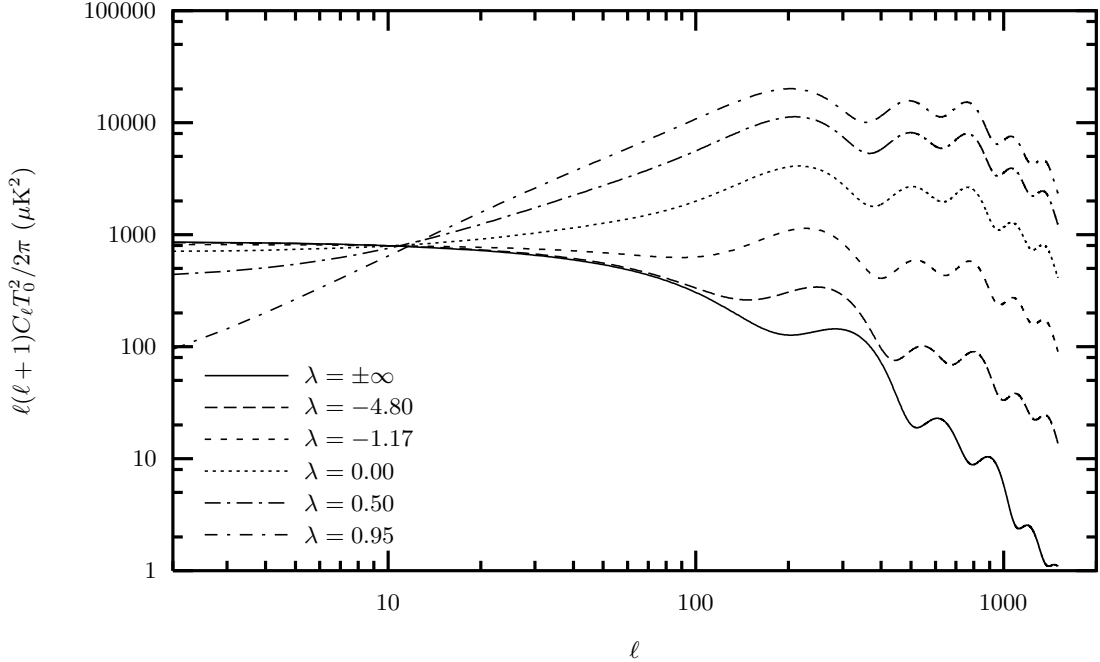 llx=0 lly=0 urx=600 ury=420 rwi=7200}
\put(963,413){\makebox(0,0)[l]{$\lambda = 0.95$}}%
\put(963,513){\makebox(0,0)[l]{$\lambda = 0.50$}}%
\put(963,613){\makebox(0,0)[l]{$\lambda = 0.00$}}%
\put(963,713){\makebox(0,0)[l]{$\lambda = -1.17$}}%
\put(963,813){\makebox(0,0)[l]{$\lambda = -4.80$}}%
\put(963,913){\makebox(0,0)[l]{$\lambda = \pm\infty$}}%
\put(2000,50){\makebox(0,0){$\ell$}}%
\put(100,1180){%
\special{ps: gsave currentpoint currentpoint translate
270 rotate neg exch neg exch translate}%
\makebox(0,0)[b]{\shortstack{$\ell(\ell+1) C_\ell T_0^2 / 2 \pi$ ($\mu$K$^2$)}}%
\special{ps: currentpoint grestore moveto}%
}%
\put(3159,200){\makebox(0,0){1000}}%
\put(2192,200){\makebox(0,0){100}}%
\put(1226,200){\makebox(0,0){10}}%
\put(500,2060){\makebox(0,0)[r]{100000}}%
\put(500,1708){\makebox(0,0)[r]{10000}}%
\put(500,1356){\makebox(0,0)[r]{1000}}%
\put(500,1004){\makebox(0,0)[r]{100}}%
\put(500,652){\makebox(0,0)[r]{10}}%
\put(500,300){\makebox(0,0)[r]{1}}%
\end{picture}%
\endgroup
 
\caption{CMB anisotropies in CDM-type correlated hybrid models for
various values of the parameter $\lambda$. The highest curve is
studied in more details in \FIG{\ref{fig_dop}}, the dotted curve
represents the (standard) adiabatic case, and the lowest represents
the pure isocurvature case shown in \FIG{\ref{fig_iso}}. Note that the
height of the acoustic peaks with respect to the Sachs-Wolfe plateau
varies with $\lambda$, according to
\EQ{\ref{root_lambda}}.}
\label{fig_cor1dt}
\end{figure}

\begin{figure}
\centering
% GNUPLOT: LaTeX picture with Postscript
\begingroup%
  \makeatletter%
  \newcommand{\GNUPLOTspecial}{%
    \@sanitize\catcode`\%=14\relax\special}%
  \setlength{\unitlength}{0.12bp}%
\begin{picture}(3600,2160)(0,0)%
\special{psfile=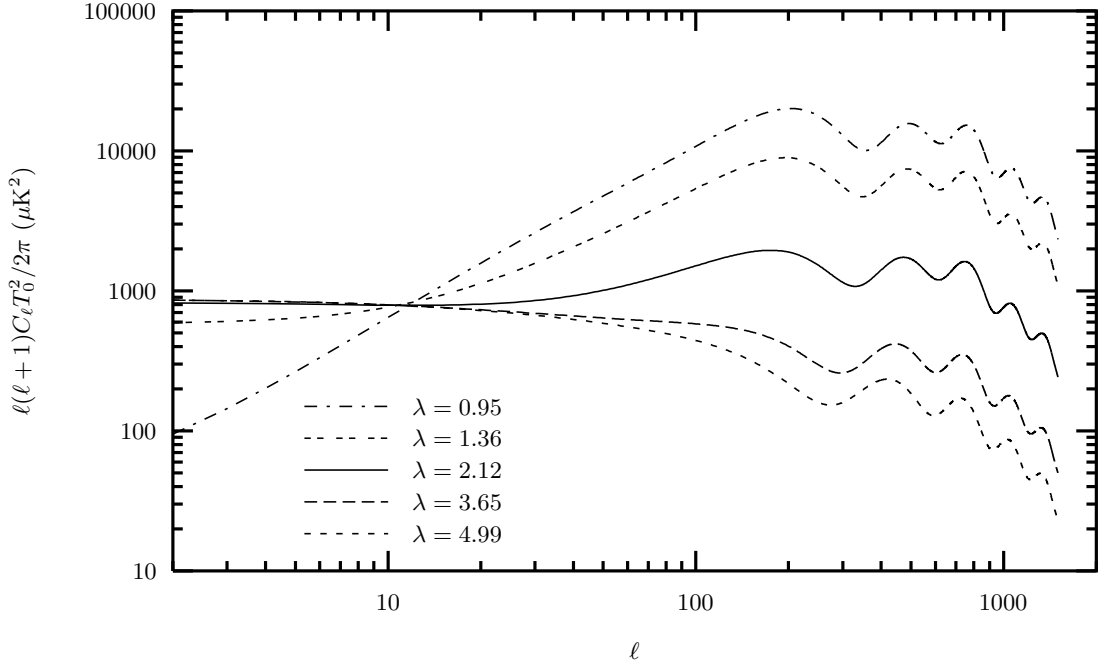 llx=0 lly=0 urx=600 ury=420 rwi=7200}
\put(1276,417){\makebox(0,0)[l]{ $\lambda = 4.99$ }}%
\put(1276,517){\makebox(0,0)[l]{ $\lambda = 3.65$ }}%
\put(1276,617){\makebox(0,0)[l]{ $\lambda = 2.12$ }}%
\put(1276,717){\makebox(0,0)[l]{ $\lambda = 1.36$ }}%
\put(1276,817){\makebox(0,0)[l]{ $\lambda = 0.95$ }}%
\put(2000,50){\makebox(0,0){$\ell$}}%
\put(100,1180){%
\special{ps: gsave currentpoint currentpoint translate
270 rotate neg exch neg exch translate}%
\makebox(0,0)[b]{\shortstack{$\ell(\ell+1) C_\ell T_0^2 / 2 \pi$ ($\mu$K$^2$)}}%
\special{ps: currentpoint grestore moveto}%
}%
\put(3159,200){\makebox(0,0){1000}}%
\put(2192,200){\makebox(0,0){100}}%
\put(1226,200){\makebox(0,0){10}}%
\put(500,2060){\makebox(0,0)[r]{100000}}%
\put(500,1620){\makebox(0,0)[r]{10000}}%
\put(500,1180){\makebox(0,0)[r]{1000}}%
\put(500,740){\makebox(0,0)[r]{100}}%
\put(500,300){\makebox(0,0)[r]{10}}%
\end{picture}%
\endgroup
 
\caption{CMB anisotropies in  CDM-type correlated hybrid models for
various values of the parameter $\lambda$. The two highest curves are
studied in more details in \FIGS{\ref{fig_isw}} and
\ref{fig_dop}. }
\label{fig_cor2dt}
\end{figure}

\begin{figure}
\centering
% GNUPLOT: LaTeX picture with Postscript
\begingroup%
  \makeatletter%
  \newcommand{\GNUPLOTspecial}{%
    \@sanitize\catcode`\%=14\relax\special}%
  \setlength{\unitlength}{0.12bp}%
\begin{picture}(3600,2160)(0,0)%
\special{psfile=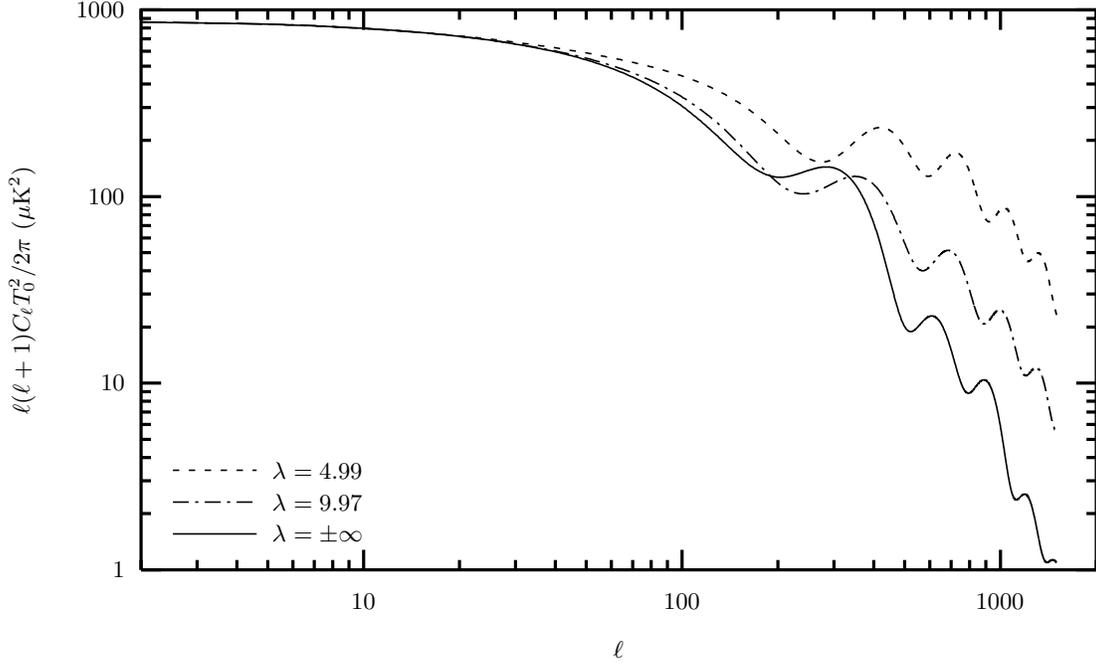 llx=0 lly=0 urx=600 ury=420 rwi=7200}
\put(863,413){\makebox(0,0)[l]{$\lambda = \pm\infty$}}%
\put(863,513){\makebox(0,0)[l]{$\lambda = 9.97$}}%
\put(863,613){\makebox(0,0)[l]{$\lambda = 4.99$}}%
\put(1950,50){\makebox(0,0){$\ell$}}%
\put(100,1180){%
\special{ps: gsave currentpoint currentpoint translate
270 rotate neg exch neg exch translate}%
\makebox(0,0)[b]{\shortstack{$\ell(\ell+1) C_\ell T_0^2 / 2 \pi$ ($\mu$K$^2$)}}%
\special{ps: currentpoint grestore moveto}%
}%
\put(3149,200){\makebox(0,0){1000}}%
\put(2149,200){\makebox(0,0){100}}%
\put(1149,200){\makebox(0,0){10}}%
\put(400,2060){\makebox(0,0)[r]{1000}}%
\put(400,1473){\makebox(0,0)[r]{100}}%
\put(400,887){\makebox(0,0)[r]{10}}%
\put(400,300){\makebox(0,0)[r]{1}}%
\end{picture}%
\endgroup
 
\caption{CMB anisotropies in CDM-type correlated hybrid models for large
(positive) values of the parameter $\lambda$. The solid curve represents
the pure isocurvature case of \FIG{\ref{fig_iso}}. Note that the
height of the acoustic  peaks with respect to the Sachs-Wolfe plateau
varies slowly in this  range of values for $\lambda$.}
\label{fig_cor3dt}
\end{figure}

In the case of hybrid perturbations, we will be somehow between these
two extreme situations. For convenience, let us parametrize the hybrid
perturbations by $\lambda$, which is defined in the relation
\beq
S=\lambda \hat \Phi,
\eeq
which will quantify how far we are from a purely adiabatic model. The
case $\lambda=0$ corresponds to pure adiabatic initial conditions,
whereas the limit where $\lambda$ goes to infinity corresponds to pure
isocurvature perturbations.  $\lambda$ can be positive or negative. To
be more specific, one can call the hybrid perturbations we are
studying {\it correlated} when $\lambda >0$ and {\it anticorrelated}
when $\lambda <0$.  In \FIGS{\ref{fig_cor1dt}-\ref{fig_cor3dt}}, we
have plotted the total temperature anisotropy as a function of the
multipole index $\ell$, for various values of the parameter $\lambda$
(for CDM-type hybrid perturbations), and keeping the same
normalization at large angular scales.

\begin{figure}
\centering
% GNUPLOT: LaTeX picture with Postscript
\begingroup%
  \makeatletter%
  \newcommand{\GNUPLOTspecial}{%
    \@sanitize\catcode`\%=14\relax\special}%
  \setlength{\unitlength}{0.12bp}%
\begin{picture}(3600,2160)(0,0)%
\special{psfile=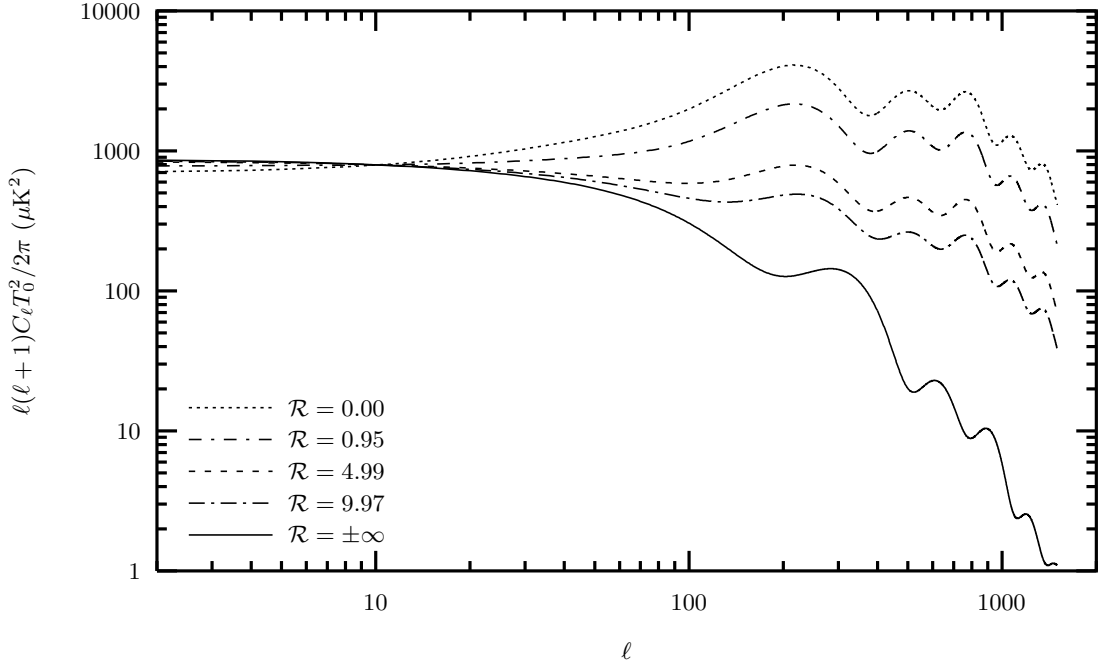 llx=0 lly=0 urx=600 ury=420 rwi=7200}
\put(913,413){\makebox(0,0)[l]{${\cal R} = \pm\infty$}}%
\put(913,513){\makebox(0,0)[l]{${\cal R} = 9.97$}}%
\put(913,613){\makebox(0,0)[l]{${\cal R} = 4.99$}}%
\put(913,713){\makebox(0,0)[l]{${\cal R} = 0.95$}}%
\put(913,813){\makebox(0,0)[l]{${\cal R} = 0.00$}}%
\put(1975,50){\makebox(0,0){$\ell$}}%
\put(100,1180){%
\special{ps: gsave currentpoint currentpoint translate
270 rotate neg exch neg exch translate}%
\makebox(0,0)[b]{\shortstack{$\ell(\ell+1) C_\ell T_0^2 / 2 \pi$ ($\mu$K$^2$)}}%
\special{ps: currentpoint grestore moveto}%
}%
\put(3154,200){\makebox(0,0){1000}}%
\put(2171,200){\makebox(0,0){100}}%
\put(1187,200){\makebox(0,0){10}}%
\put(450,2060){\makebox(0,0)[r]{10000}}%
\put(450,1620){\makebox(0,0)[r]{1000}}%
\put(450,1180){\makebox(0,0)[r]{100}}%
\put(450,740){\makebox(0,0)[r]{10}}%
\put(450,300){\makebox(0,0)[r]{1}}%
\end{picture}%
\endgroup
 
\caption{CMB anisotropies in independant (uncorrelated) hybrid
CDM-type models for various values of the parameter ${\cal R}$. The
lowest, solid curve represents the pure isocurvature case of
\FIG{\ref{fig_iso}} (${\cal R} = \infty$), and the highest, dotted curve
represents the adiabatic case (${\cal R} = 0$). We have chosen for
${\cal R}$ the same numerical values as for $\lambda$ in
\FIGS{\ref{fig_cor1dt}-\ref{fig_cor3dt}}. Note the significant
difference between the correlated and the independant cases especially
in the region where ${\cal R},\lambda \simeq 1$.}
\label{fig_ncordt}
\end{figure}

To emphasize the difference between correlated hybrid perturbations
and independent hybrid perturbations, which have been considered in
the literature, we have plotted , in \FIG{\ref{fig_ncordt}}, the total
temperature anisotropy for independent hybrid initial conditions. The
curves are parametrized by the number ${\cal R}$, which is defined by
\beq
{\cal P}_S^{1/2}={\cal R} {\cal P}_\Phi^{1/2}. 
\eeq
In some sense ${\cal R}$ is the analogous, in the independent case, of
$\lambda$ since the square of both quantities corresponds to the ratio
of the power spectra. But, of course ${\cal R}$ can be only
positive. The way these curves are obtained is also different. Whereas
for the correlated mixtures, one implements hybrid initial conditions
from the beginning and one runs the Boltzmann code once (per model),
in the case of independent mixtures, one runs the code first with
purely adiabatic initial conditions, then a second time with purely
isocurvature initial conditions, and the final $C_\ell$ are obtained
by a weighted sum of the $C_\ell$ obtained from each run. As a
consequence, the first acoustic peak, as well as the subsequent ones,
will always appear lower, relatively to the plateau in the hybrid case
than in the purely adiabatic case.

The behaviour of the $C_\ell$ for the correlated hybrid models is
quite different when one increases the isocurvature proportion.  For
anticorrelated perturbations, \IE{} $\lambda <0$, the behaviour is
what is expected naively~: the amplitude of the peaks decreases,
relatively to the plateau, with higher proportion of isocurvature
perturbations, as illustrated in \FIG{\ref{fig_cor1dt}} with the
curves lower than the adiabatic case.  But the evolution is more
complicated when one considers correlated models, \IE{} with
$\lambda>0$. Starting from the adiabatic case ($\lambda=0$) and
increasing slowly $\lambda$, one begins with a phase where the
amplitude of the peaks will increase with respect to the plateau, as
illustrated by the curves above the adiabatic one in
\FIG{\ref{fig_cor1dt}}. If one goes on increasing $\lambda$, one
reaches a critical value, beyond which the peaks will now diminish
with increasing $\lambda$, as illustrated by the curves of
\FIG{\ref{fig_cor2dt}}.

\begin{figure}
\centering
% GNUPLOT: LaTeX picture with Postscript
\begingroup%
  \makeatletter%
  \newcommand{\GNUPLOTspecial}{%
    \@sanitize\catcode`\%=14\relax\special}%
  \setlength{\unitlength}{0.12bp}%
\begin{picture}(3600,2160)(0,0)%
\special{psfile=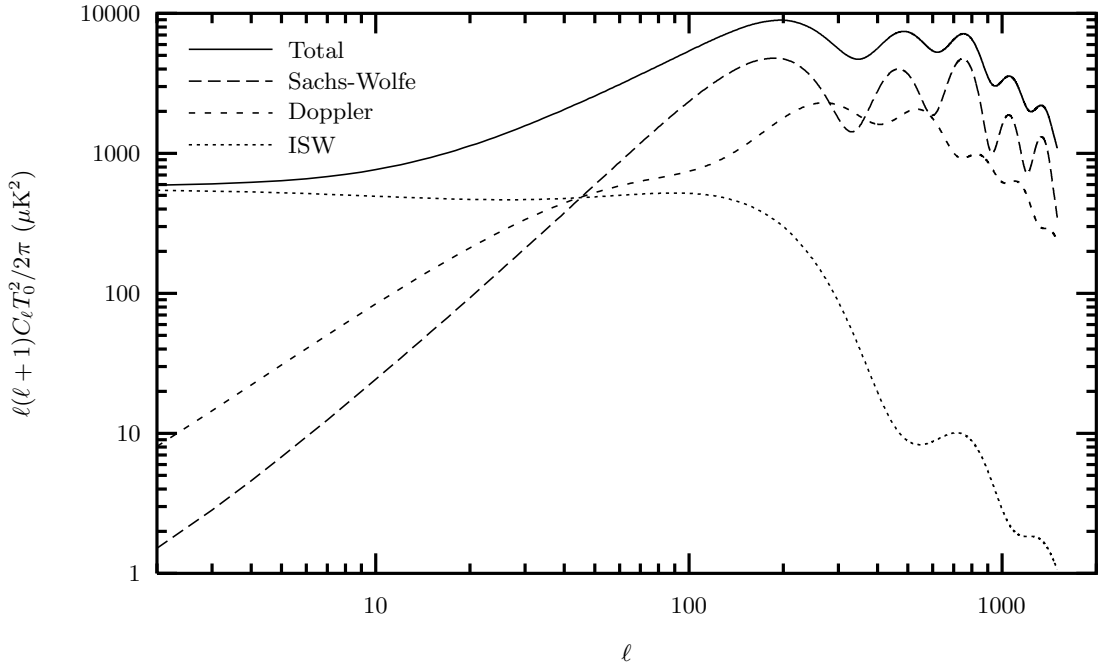 llx=0 lly=0 urx=600 ury=420 rwi=7200}
\put(913,1647){\makebox(0,0)[l]{ISW}}%
\put(913,1747){\makebox(0,0)[l]{Doppler}}%
\put(913,1847){\makebox(0,0)[l]{Sachs-Wolfe}}%
\put(913,1947){\makebox(0,0)[l]{Total}}%
\put(1975,50){\makebox(0,0){$\ell$}}%
\put(100,1180){%
\special{ps: gsave currentpoint currentpoint translate
270 rotate neg exch neg exch translate}%
\makebox(0,0)[b]{\shortstack{$\ell(\ell+1) C_\ell T_0^2 / 2 \pi$ ($\mu$K$^2$)}}%
\special{ps: currentpoint grestore moveto}%
}%
\put(3154,200){\makebox(0,0){1000}}%
\put(2171,200){\makebox(0,0){100}}%
\put(1187,200){\makebox(0,0){10}}%
\put(450,2060){\makebox(0,0)[r]{10000}}%
\put(450,1620){\makebox(0,0)[r]{1000}}%
\put(450,1180){\makebox(0,0)[r]{100}}%
\put(450,740){\makebox(0,0)[r]{10}}%
\put(450,300){\makebox(0,0)[r]{1}}%
\end{picture}%
\endgroup
 
\caption{CMB anisotropies in a  CDM-type correlated hybrid 
model (with $\lambda = 1.36$). The solid line represents the total
(scalar) contribution. The Sachs-Wolfe (SW), Doppler and Integrated
Sachs-Wolfe contributions are respectively represented in long-dashed,
short-dashed, and dotted lines. The parameter $\lambda$ has been
chosen so that the SW contribution cancels at lowest order. In this
case, the low multipole power is no longer dominated by the SW
contribution, but rather by the ISW contribution. }
\label{fig_isw}
\end{figure}

\begin{figure}
\centering
% GNUPLOT: LaTeX picture with Postscript
\begingroup%
  \makeatletter%
  \newcommand{\GNUPLOTspecial}{%
    \@sanitize\catcode`\%=14\relax\special}%
  \setlength{\unitlength}{0.12bp}%
\begin{picture}(3600,2160)(0,0)%
\special{psfile=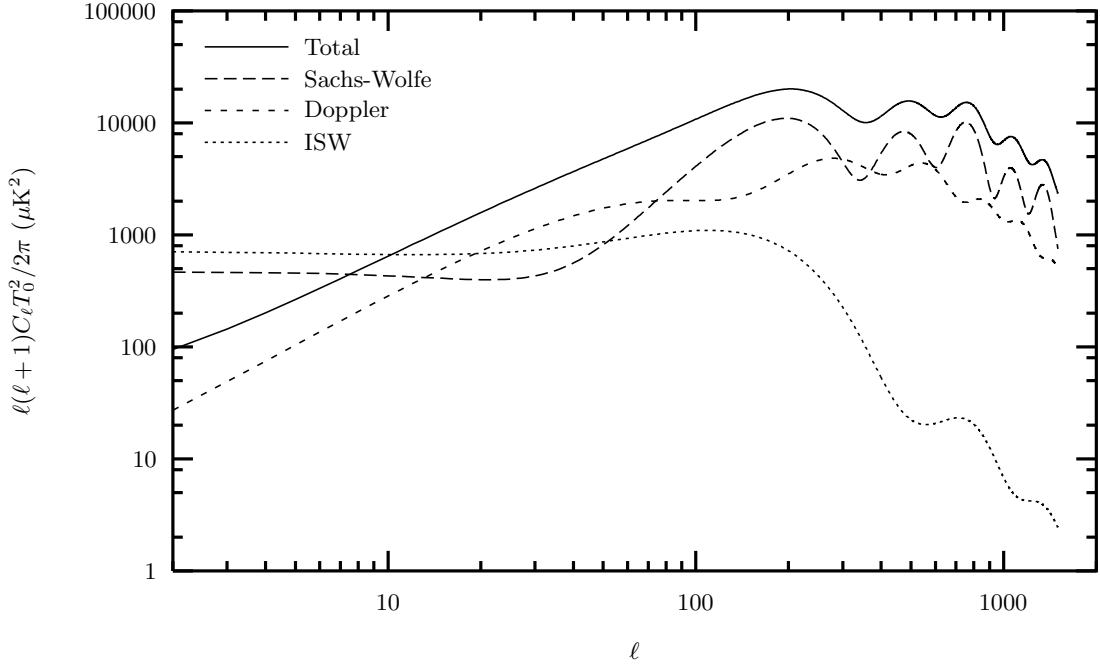 llx=0 lly=0 urx=600 ury=420 rwi=7200}
\put(963,1647){\makebox(0,0)[l]{ISW}}%
\put(963,1747){\makebox(0,0)[l]{Doppler}}%
\put(963,1847){\makebox(0,0)[l]{Sachs-Wolfe}}%
\put(963,1947){\makebox(0,0)[l]{Total}}%
\put(2000,50){\makebox(0,0){$\ell$}}%
\put(100,1180){%
\special{ps: gsave currentpoint currentpoint translate
270 rotate neg exch neg exch translate}%
\makebox(0,0)[b]{\shortstack{$\ell(\ell+1) C_\ell T_0^2 / 2 \pi$ ($\mu$K$^2$)}}%
\special{ps: currentpoint grestore moveto}%
}%
\put(3159,200){\makebox(0,0){1000}}%
\put(2192,200){\makebox(0,0){100}}%
\put(1226,200){\makebox(0,0){10}}%
\put(500,2060){\makebox(0,0)[r]{100000}}%
\put(500,1708){\makebox(0,0)[r]{10000}}%
\put(500,1356){\makebox(0,0)[r]{1000}}%
\put(500,1004){\makebox(0,0)[r]{100}}%
\put(500,652){\makebox(0,0)[r]{10}}%
\put(500,300){\makebox(0,0)[r]{1}}%
\end{picture}%
\endgroup
 
\caption{CMB anisotropies in a CDM-type correlated hybrid model (with
$\lambda = 0.95$). The solid line represent the total
contribution. The Sachs-Wolfe (SW), Doppler and Integrated Sachs-Wolfe
(ISW) contributions are respectively represented in long-dashed,
short-dashed, and dotted lines. The parameter $\lambda$ has been
chosen so that the SW and ISW contribution almost cancel each other at
lowest order. In this case, the low multipole power is dominated by
the Doppler contribution, which is not flat at low multipoles.}
\label{fig_dop}
\end{figure}

One can understand this surprising behaviour if one goes back to the
results of the previous section. In the case of CDM hybrid
perturbations, one can evaluate the SW plateau, using
\EQNS{\ref{T_adiab}} and (\ref{T_iso_c})~:
\beq
\left(\frac{\Delta T}{T}\right)_{{\rm SW}}=\left[
\frac{1}{10}\left(3+\frac{4}{5}\Omn\rd\right)
-\frac{2}{5}\Omc\md \lambda\right]\hat\Phi,
\label{sw_cdm}
\eeq
and therefore, there is indeed a critical value for $\lambda$ for
which the SW plateau is suppressed, which explains the relative height
of the peaks. In fact, things are slightly more complicated, because
when the SW term is suppressed, because this special choice of initial
conditions, the other terms which contribute to the anisotropies
cannot be neglected any longer. \FFIG{\ref{fig_isw}} shows in
particular that the plateau can be due essentially to the ISW
effect. Note that the value $\lambda=1.36$ for which this effect was
numerically obtained is slightly different from the value one would
deduce from (\ref{sw_cdm}). This is because (\ref{sw_cdm}) was
obtained by supposing that the last scattering surface is completely
in the matter-dominated era, which is not the case since the
radiation-to-matter transition occurs no very long before
recombination.  One can also adjust the initial conditions so that the
large scale anisotropies will be dominated by the Doppler term, in
which case there is no longer a plateau on large scales, but an
increasing slope as can be seen on \FIG{\ref{fig_dop}}.

\begin{figure}
\centering
% GNUPLOT: LaTeX picture with Postscript
\begingroup%
  \makeatletter%
  \newcommand{\GNUPLOTspecial}{%
    \@sanitize\catcode`\%=14\relax\special}%
  \setlength{\unitlength}{0.12bp}%
\begin{picture}(3600,2160)(0,0)%
\special{psfile=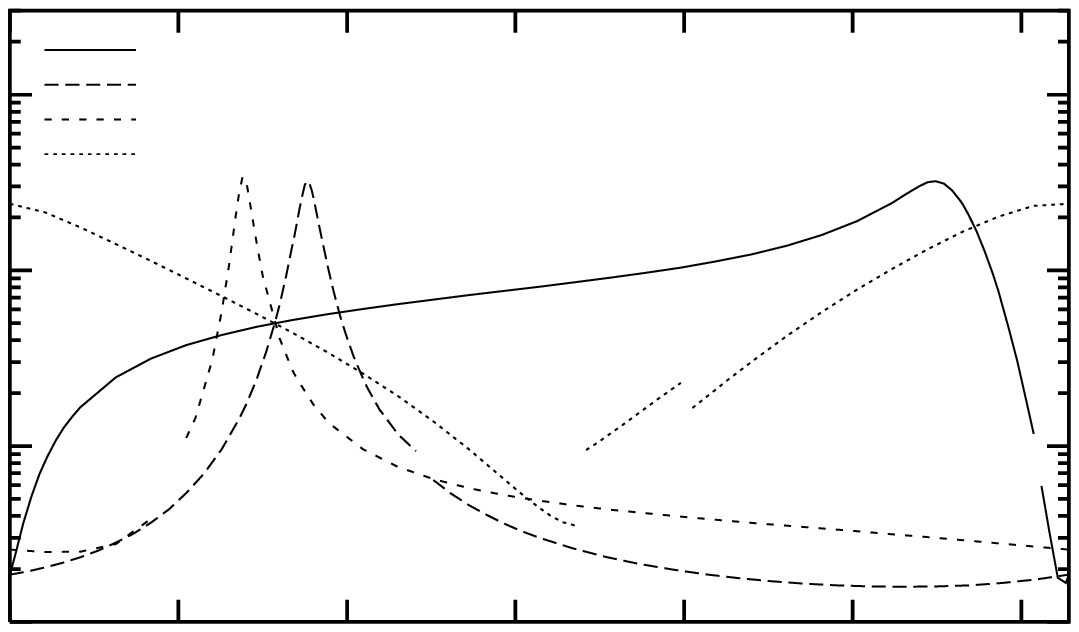 llx=0 lly=0 urx=600 ury=420 rwi=7200}
\put(813,1647){\makebox(0,0)[l]{Isocurvature neutrinos}}%
\put(813,1747){\makebox(0,0)[l]{Isocurvature photons}}%
\put(813,1847){\makebox(0,0)[l]{Isocurvature CDM}}%
\put(813,1947){\makebox(0,0)[l]{Isocurvature baryons}}%
\put(1925,50){\makebox(0,0){ $\theta$ }}%
\put(100,1180){%
\special{ps: gsave currentpoint currentpoint translate
270 rotate neg exch neg exch translate}%
\makebox(0,0)[b]{\shortstack{Ratio 1$^{\rm st}$ peak height / SW plateau}}%
\special{ps: currentpoint grestore moveto}%
}%
\put(3313,200){\makebox(0,0){3}}%
\put(2827,200){\makebox(0,0){2.5}}%
\put(2342,200){\makebox(0,0){2}}%
\put(1856,200){\makebox(0,0){1.5}}%
\put(1371,200){\makebox(0,0){1}}%
\put(885,200){\makebox(0,0){0.5}}%
\put(400,200){\makebox(0,0){0}}%
\put(350,1818){\makebox(0,0)[r]{100}}%
\put(350,1312){\makebox(0,0)[r]{10}}%
\put(350,806){\makebox(0,0)[r]{1}}%
\put(350,300){\makebox(0,0)[r]{0.1}}%
\end{picture}%
\endgroup
 
\caption{Ratio of the height of the first acoustic peak to the height
of the Sachs-Wolfe (SW) plateau for the four types of hybrid
models. The height of the SW plateau is obtained by averaging the
power between $\ell = 2$ and $\ell = 25$, which correspond to the
angular scales explored by COBE. We have defined the ``first acoustic
peak''as the first maximum of the multipole anisotropy spectrum for
$\ell \geq 100$.  Each ratio peaks at a high value ($\simeq 30$) which
roughly corresponds to the moment where the SW plateau disappears [see
\EQ{\ref{root_lambda}}; in practice, this occurs when the SW and ISW
contributions cancel at low multipoles]. The four curves intersect at
$\theta = \pi/4$, as expected, since this value corresponds to the pure
adiabatic case [see \EQ{\ref{def_theta}}].  As $\theta$ varies, the
position on the $\ell$-axis of the first acoustic peak slightly shifts
to the left or to the right (see also \FIG{\ref{fig_pic_pos}}). In
some cases, the peak goes below $\ell = 100$. In this case the ``new''
first acoustic peak position is at $\ell \simeq 200-300$, and its
height is different, hence the discontinuities in the  curves.}
\label{fig_pic_plat}
\end{figure}

Although at this stage we have discussed and illustrated only the CDM
correlated hybrid case, a similar behaviour appears for the three
other types of correlated hybrid perturbations, but with noticable
differences. We have systematically explored the parameter space for
the four types of correlated hybrid initial conditions and measured
the predicted height of the first acoustic peak with respect to the
plateau.  The results are given in \FIG{\ref{fig_pic_plat}}. Here, we
have adopted a different parametrization of the hybrid correlated
perturbations so that one can represent easily all cases. We have
defined an angular variable $\theta_X$ so that our initial conditions
for the density contrasts are of the form
\beq
{\D_X\over 1+w_X}=\cos\theta_X,\qquad {\D_A\over 1+w_A}=\sin\theta_X, 
\quad A\neq X.
\eeq
which implies
\beq
S_X=(\cot\theta_X-1)\Da.
\label{def_theta}
\eeq
Of course, this parameter $\theta_X$ can be related to the parameter
$\lambda$.  Let us write
\beq
\Da=\alpha\hat\Phi+\beta_X S_X,
\eeq
where the subscript $X$ for the coefficient $\beta_X$ refers to the
type of hybrid mode we are considering. $\alpha$ is the same for all
four types and is given by the coefficient in \EQ{\ref{Da}}. $\beta_X$
is zero for the CDM and baryon modes and is given by the second term
on the right hand side in \EQNS{\ref{Da_p}} and (\ref{Da_n}) for the
photon and neutrino cases respectively.  The relation between
$\lambda_X$ and $\theta_X$ is then
\beq
\label{theta_lambda}
\lambda_X={(\cot\theta_X-1)\alpha\over 1- (\cot\theta_X-1)\beta_X}.
\eeq
In the region near $\theta=\pi/4$ (corresponding to the pure adiabatic
case and where all curves cross), it is easy to see that the relation
between $\theta$ and $\lambda$ is the same for the four types of
perturbations and is given numerically by
\beq
\lambda \simeq 2 \alpha (\theta-\pi/4) \simeq 9.97 (\theta-\pi/4).
\eeq

Let us see what happens when one deviates from the pure adiabatic
case. In the baryon case, the first peak will increase for correlated
mixtures and decrease for anticorrelated mixtures, as in the CDM case,
although in a much more moderate way. On the contrary, in the photon
and neutrino cases, the first peak will start to increase for {\it
anticorrelated} perturbations. Moreover, the evolution in the neutrino
case is slower than in the photon case.

All these results can be understood rather easily with the analytical
results of the previous section. Indeed, in all cases, the SW
anisotropy term can be written in the form
\beq
\label{root_lambda}
\left(\frac{\Delta T}{T}\right)_{{\rm SW}}=\left(a+b_X\lambda_X\right)
\hat\Phi,
\eeq
where the special case of CDM is given just above. The coefficient $a$
is the same for all four cases, as was shown in the previous
section. Therefore, the evolution of the SW plateau will be determined
by the coefficient $b_X$ which is different in each case. For baryon
and CDM perturbations, the coefficient $b$ is negative, which explains
why the increase of the first peak (or equivalently the decrease of
the plateau) corresponds to correlated ($\lambda>0$) perturbations,
when one deviates from the pure adiabatic case. Moreover,
$|b_c|>|b_b|$ because $\Omega_c>\Omega_b$ and therefore the response
to the increase of the isocurvature proportion is stronger in the CDM
case. In the photon and neutrino cases, the coefficient $\beta$ is
positive and therefore only anticorrelated perturbations can lead to
an increase of the peak. Moreover, $b_\gamma>b_\nu$, and similarly to
the heavy species, the response to the increase of $\lambda$ is
stronger in the photon case than in the neutrino case.  A consequence
of these results is that, potentially, the correlated hybrid
perturbations in the CDM and photon cases will be more strongly
constrained by the CMB measurements than the baryon and neutrino
cases.

\begin{figure}
\centering
% GNUPLOT: LaTeX picture with Postscript
\begingroup%
  \makeatletter%
  \newcommand{\GNUPLOTspecial}{%
    \@sanitize\catcode`\%=14\relax\special}%
  \setlength{\unitlength}{0.12bp}%
\begin{picture}(3600,2160)(0,0)%
\special{psfile=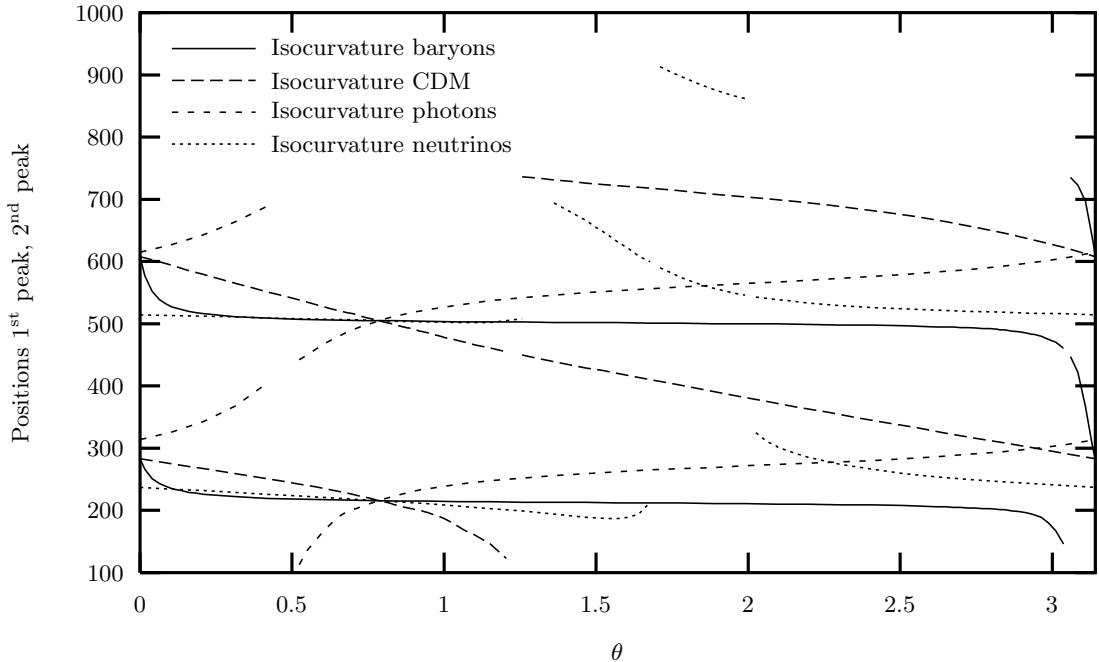 llx=0 lly=0 urx=600 ury=420 rwi=7200}
\put(863,1647){\makebox(0,0)[l]{Isocurvature neutrinos}}%
\put(863,1747){\makebox(0,0)[l]{Isocurvature photons}}%
\put(863,1847){\makebox(0,0)[l]{Isocurvature CDM}}%
\put(863,1947){\makebox(0,0)[l]{Isocurvature baryons}}%
\put(1950,50){\makebox(0,0){ $\theta$ }}%
\put(100,1180){%
\special{ps: gsave currentpoint currentpoint translate
270 rotate neg exch neg exch translate}%
\makebox(0,0)[b]{\shortstack{Positions 1$^{\rm st}$ peak, 2$^{\rm nd}$ peak}}%
\special{ps: currentpoint grestore moveto}%
}%
\put(3315,200){\makebox(0,0){3}}%
\put(2837,200){\makebox(0,0){2.5}}%
\put(2360,200){\makebox(0,0){2}}%
\put(1882,200){\makebox(0,0){1.5}}%
\put(1405,200){\makebox(0,0){1}}%
\put(927,200){\makebox(0,0){0.5}}%
\put(450,200){\makebox(0,0){0}}%
\put(400,2060){\makebox(0,0)[r]{1000}}%
\put(400,1864){\makebox(0,0)[r]{900}}%
\put(400,1669){\makebox(0,0)[r]{800}}%
\put(400,1473){\makebox(0,0)[r]{700}}%
\put(400,1278){\makebox(0,0)[r]{600}}%
\put(400,1082){\makebox(0,0)[r]{500}}%
\put(400,887){\makebox(0,0)[r]{400}}%
\put(400,691){\makebox(0,0)[r]{300}}%
\put(400,496){\makebox(0,0)[r]{200}}%
\put(400,300){\makebox(0,0)[r]{100}}%
\end{picture}%
\endgroup
 
\caption{Position of the two first acoustic peaks in the four types of
hybrid models.  As explained in \FIG{\ref{fig_pic_plat}}, the first
acoustic peak is defined as the the first maximum of the multipole
anisotropy spectrum for $\ell \geq 100$. In some cases ($\theta\simeq
3$ for the isocurvature baryon model, $\theta\simeq 0.5$ for the
isocurvature photon model and $\theta\simeq 1.2$ for the isocurvature
CDM model), the first peak disappears in the low multipole region, and
the ``new'' first peak becomes the ``former'' second one. In some
other cases, the peaks is smeared in the power spectrum and
disappears, as it is the case for the first acoustic peak in the
isocurvature neutrino model at $\theta\simeq 1.2$, and at
$\theta\simeq 1.7$ for the second one.}
\label{fig_pic_pos}
\end{figure}

It is also important, in the spirit of putting constraints on this
type of modes, to see the position of the two Doppler peaks on the
$\ell$-axis. We have plotted in \FIG{\ref{fig_pic_pos}}, the positions
of the first and second acoustic peaks. Near the pure adiabatic case,
the behaviour is once more strongly pronounced in the CDM and photon
cases. In this region, in the CDM case, correlated hybrid
perturbations tend to displace the peaks to smaller $\ell$ whereas
anticorrelated perturbations push the peaks to higher $\ell$. The same
behaviour, but very attenuated, seems to apply to the baryon and
neutrino cases. In contrast, in the photon case, higher $\ell$
correspond to correlated perturbations and lower $\ell$ to
anticorrelated perturbations.

\begin{figure}
\centering
% GNUPLOT: LaTeX picture with Postscript
\begingroup%
  \makeatletter%
  \newcommand{\GNUPLOTspecial}{%
    \@sanitize\catcode`\%=14\relax\special}%
  \setlength{\unitlength}{0.12bp}%
\begin{picture}(3600,2160)(0,0)%
\special{psfile=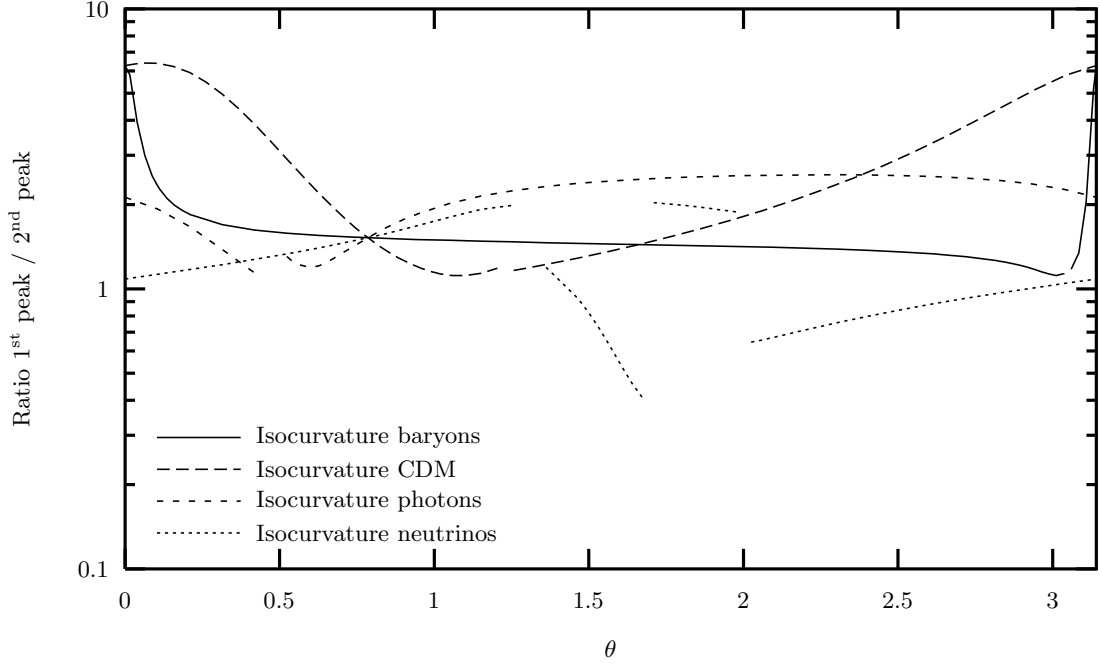 llx=0 lly=0 urx=600 ury=420 rwi=7200}
\put(813,413){\makebox(0,0)[l]{Isocurvature neutrinos}}%
\put(813,513){\makebox(0,0)[l]{Isocurvature photons}}%
\put(813,613){\makebox(0,0)[l]{Isocurvature CDM}}%
\put(813,713){\makebox(0,0)[l]{Isocurvature baryons}}%
\put(1925,50){\makebox(0,0){ $\theta$ }}%
\put(100,1180){%
\special{ps: gsave currentpoint currentpoint translate
270 rotate neg exch neg exch translate}%
\makebox(0,0)[b]{\shortstack{Ratio 1$^{\rm st}$ peak / 2$^{\rm nd}$ peak}}%
\special{ps: currentpoint grestore moveto}%
}%
\put(3313,200){\makebox(0,0){3}}%
\put(2827,200){\makebox(0,0){2.5}}%
\put(2342,200){\makebox(0,0){2}}%
\put(1856,200){\makebox(0,0){1.5}}%
\put(1371,200){\makebox(0,0){1}}%
\put(885,200){\makebox(0,0){0.5}}%
\put(400,200){\makebox(0,0){0}}%
\put(350,2060){\makebox(0,0)[r]{10}}%
\put(350,1180){\makebox(0,0)[r]{1}}%
\put(350,300){\makebox(0,0)[r]{0.1}}%
\end{picture}%
\endgroup
 
\caption{Relative  heights of the first two acoustic  peaks for the
four types of hybrid models. The various discontinuities of the
four curves originate from the discontinuities in the peak positions, 
explained and illustrated in \FIG{\ref{fig_pic_pos}}.}
\label{fig_pic_haut}
\end{figure}

Another important feature of the acoustic peaks is the impact of the
isocurvature part on the amplitude of the second
peak. \FFIG{\ref{fig_pic_haut}} gives the evolution of the relative
amplitude of the second peak with respect to the first peak, when one
varies the coefficient $\lambda$, for the four types of
perturbations. As before, in the vicinity of the pure adiabatic point,
the effects are more important in the CDM and photons cases than in
the two other cases.

\subsection{Power spectrum}

In the previous subsection, we have analysed the CMB
fluctuations. Another important part of observational data comes from
the large-scale structures.  However, it is more difficult to infer as
precise information from the LSS data as from the expected CMB
measurements, because of additional complications such as the bias
effect. Moreover, the signatures due to pure adiabatic perturbations
and pure isocurvature perturbations respectively is easier to
distinguish in the CMB anisotropies than in the matter power
spectrum. But, nevertheless, the LSS power spectrum is a useful tool,
at least to check the overall amplitude of the perturbations.

\begin{figure}
\centering
% GNUPLOT: LaTeX picture with Postscript
\begingroup%
  \makeatletter%
  \newcommand{\GNUPLOTspecial}{%
    \@sanitize\catcode`\%=14\relax\special}%
  \setlength{\unitlength}{0.12bp}%
\begin{picture}(3600,2160)(0,0)%
\special{psfile=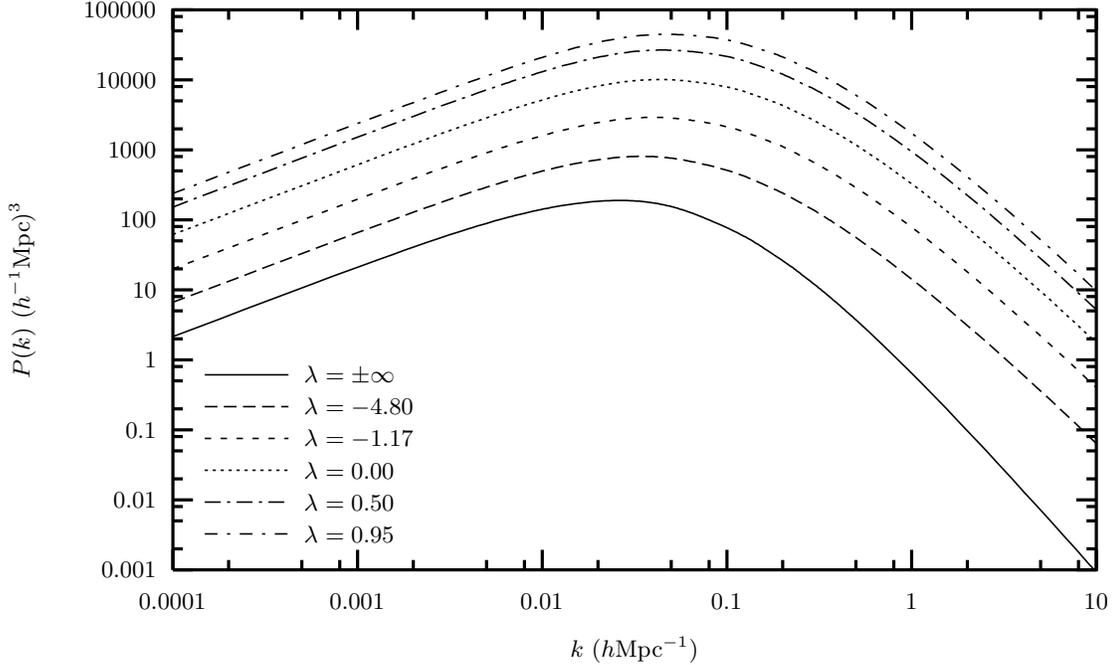 llx=0 lly=0 urx=600 ury=420 rwi=7200}
\put(963,413){\makebox(0,0)[l]{$\lambda = 0.95$}}%
\put(963,513){\makebox(0,0)[l]{$\lambda = 0.50$}}%
\put(963,613){\makebox(0,0)[l]{$\lambda = 0.00$}}%
\put(963,713){\makebox(0,0)[l]{$\lambda = -1.17$}}%
\put(963,813){\makebox(0,0)[l]{$\lambda = -4.80$}}%
\put(963,913){\makebox(0,0)[l]{$\lambda = \pm\infty$}}%
\put(2000,50){\makebox(0,0){$k$ ($h$Mpc$^{-1}$)}}%
\put(100,1180){%
\special{ps: gsave currentpoint currentpoint translate
270 rotate neg exch neg exch translate}%
\makebox(0,0)[b]{\shortstack{$P(k)$ ($h^{-1}$Mpc)$^3$}}%
\special{ps: currentpoint grestore moveto}%
}%
\put(3450,200){\makebox(0,0){10}}%
\put(2870,200){\makebox(0,0){1}}%
\put(2290,200){\makebox(0,0){0.1}}%
\put(1710,200){\makebox(0,0){0.01}}%
\put(1130,200){\makebox(0,0){0.001}}%
\put(550,200){\makebox(0,0){0.0001}}%
\put(500,2060){\makebox(0,0)[r]{100000}}%
\put(500,1840){\makebox(0,0)[r]{10000}}%
\put(500,1620){\makebox(0,0)[r]{1000}}%
\put(500,1400){\makebox(0,0)[r]{100}}%
\put(500,1180){\makebox(0,0)[r]{10}}%
\put(500,960){\makebox(0,0)[r]{1}}%
\put(500,740){\makebox(0,0)[r]{0.1}}%
\put(500,520){\makebox(0,0)[r]{0.01}}%
\put(500,300){\makebox(0,0)[r]{0.001}}%
\end{picture}%
\endgroup
 
\caption{Matter power spectra in the CDM-type hybrid models for
various values of the parameter $\lambda$. All the curves have been
normalised to COBE. Note that the overall amplitude, as well as the
position of the maximum varies with $\lambda$.}
\label{fig_cor1ps}
\end{figure}

\begin{figure}
\centering
% GNUPLOT: LaTeX picture with Postscript
\begingroup%
  \makeatletter%
  \newcommand{\GNUPLOTspecial}{%
    \@sanitize\catcode`\%=14\relax\special}%
  \setlength{\unitlength}{0.12bp}%
\begin{picture}(3600,2160)(0,0)%
\special{psfile=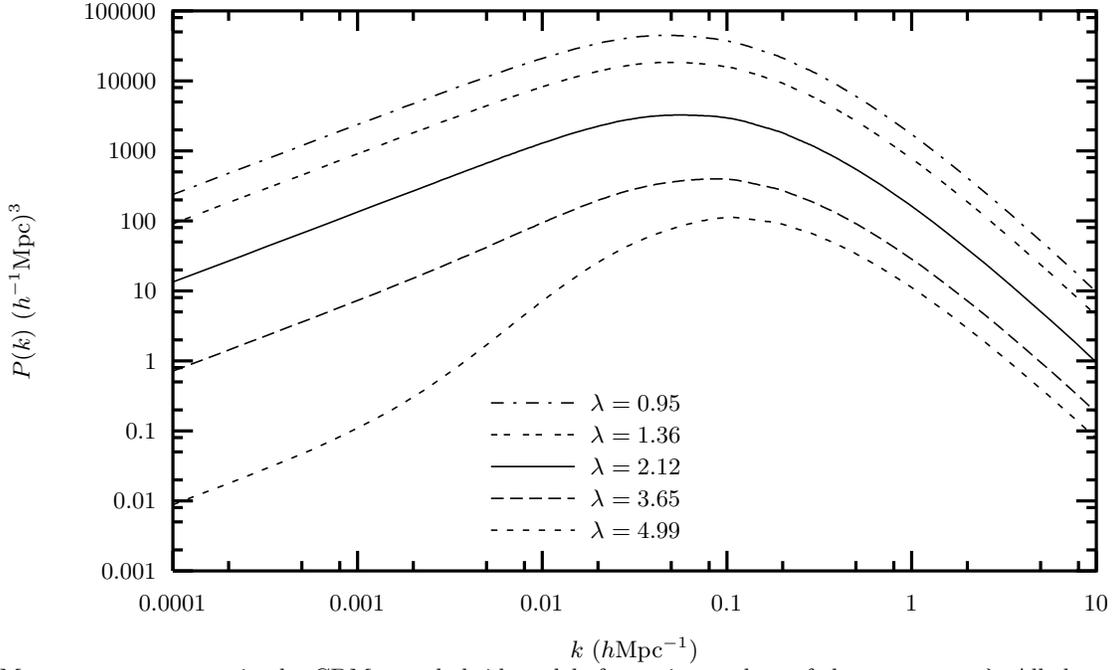 llx=0 lly=0 urx=600 ury=420 rwi=7200}
\put(1862,428){\makebox(0,0)[l]{$\lambda = 4.99$}}%
\put(1862,528){\makebox(0,0)[l]{$\lambda = 3.65$}}%
\put(1862,628){\makebox(0,0)[l]{$\lambda = 2.12$}}%
\put(1862,728){\makebox(0,0)[l]{$\lambda = 1.36$}}%
\put(1862,828){\makebox(0,0)[l]{$\lambda = 0.95$}}%
\put(2000,50){\makebox(0,0){$k$ ($h$Mpc$^{-1}$)}}%
\put(100,1180){%
\special{ps: gsave currentpoint currentpoint translate
270 rotate neg exch neg exch translate}%
\makebox(0,0)[b]{\shortstack{$P(k)$ ($h^{-1}$Mpc)$^3$}}%
\special{ps: currentpoint grestore moveto}%
}%
\put(3450,200){\makebox(0,0){10}}%
\put(2870,200){\makebox(0,0){1}}%
\put(2290,200){\makebox(0,0){0.1}}%
\put(1710,200){\makebox(0,0){0.01}}%
\put(1130,200){\makebox(0,0){0.001}}%
\put(550,200){\makebox(0,0){0.0001}}%
\put(500,2060){\makebox(0,0)[r]{100000}}%
\put(500,1840){\makebox(0,0)[r]{10000}}%
\put(500,1620){\makebox(0,0)[r]{1000}}%
\put(500,1400){\makebox(0,0)[r]{100}}%
\put(500,1180){\makebox(0,0)[r]{10}}%
\put(500,960){\makebox(0,0)[r]{1}}%
\put(500,740){\makebox(0,0)[r]{0.1}}%
\put(500,520){\makebox(0,0)[r]{0.01}}%
\put(500,300){\makebox(0,0)[r]{0.001}}%
\end{picture}%
\endgroup
 
\caption{Matter power spectra in the CDM-type hybrid models for
various values of the parameter $\lambda$. All the curves have been
normalised to COBE.  }
\label{fig_cor2ps}
\end{figure}

\begin{figure}
\centering
% GNUPLOT: LaTeX picture with Postscript
\begingroup%
  \makeatletter%
  \newcommand{\GNUPLOTspecial}{%
    \@sanitize\catcode`\%=14\relax\special}%
  \setlength{\unitlength}{0.12bp}%
\begin{picture}(3600,2160)(0,0)%
\special{psfile=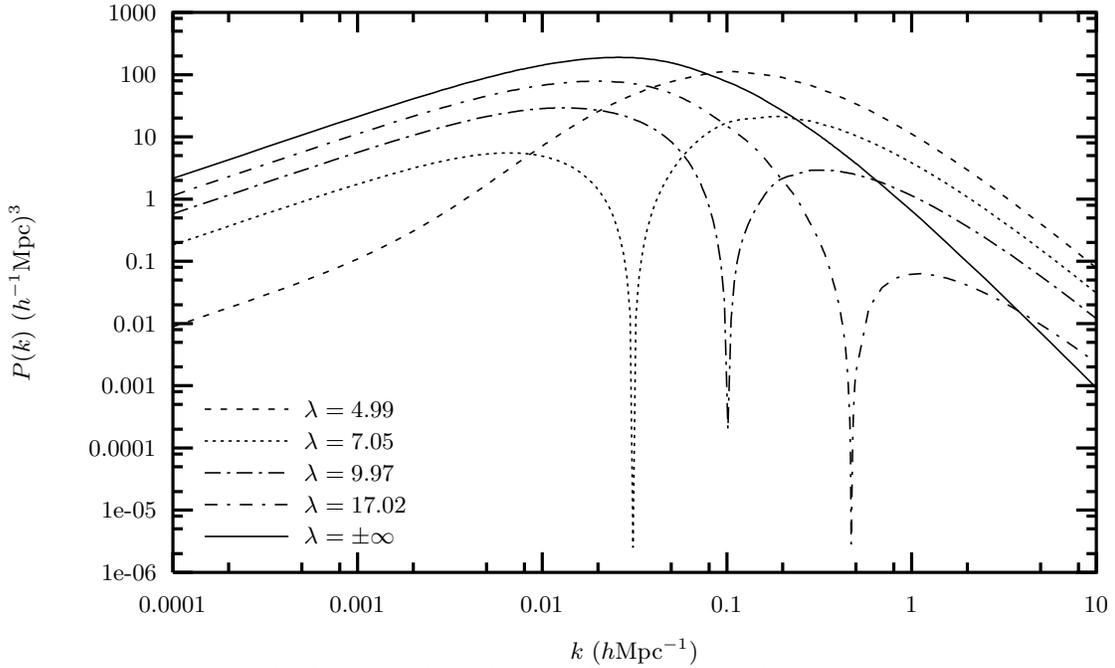 llx=0 lly=0 urx=600 ury=420 rwi=7200}
\put(963,413){\makebox(0,0)[l]{$\lambda = \pm\infty$}}%
\put(963,513){\makebox(0,0)[l]{$\lambda = 17.02$}}%
\put(963,613){\makebox(0,0)[l]{$\lambda = 9.97$}}%
\put(963,713){\makebox(0,0)[l]{$\lambda = 7.05$}}%
\put(963,813){\makebox(0,0)[l]{$\lambda = 4.99$}}%
\put(2000,50){\makebox(0,0){$k$ ($h$Mpc$^{-1}$)}}%
\put(100,1180){%
\special{ps: gsave currentpoint currentpoint translate
270 rotate neg exch neg exch translate}%
\makebox(0,0)[b]{\shortstack{$P(k)$ ($h^{-1}$Mpc)$^3$}}%
\special{ps: currentpoint grestore moveto}%
}%
\put(3450,200){\makebox(0,0){10}}%
\put(2870,200){\makebox(0,0){1}}%
\put(2290,200){\makebox(0,0){0.1}}%
\put(1710,200){\makebox(0,0){0.01}}%
\put(1130,200){\makebox(0,0){0.001}}%
\put(550,200){\makebox(0,0){0.0001}}%
\put(500,2060){\makebox(0,0)[r]{1000}}%
\put(500,1864){\makebox(0,0)[r]{100}}%
\put(500,1669){\makebox(0,0)[r]{10}}%
\put(500,1473){\makebox(0,0)[r]{1}}%
\put(500,1278){\makebox(0,0)[r]{0.1}}%
\put(500,1082){\makebox(0,0)[r]{0.01}}%
\put(500,887){\makebox(0,0)[r]{0.001}}%
\put(500,691){\makebox(0,0)[r]{0.0001}}%
\put(500,496){\makebox(0,0)[r]{1e-05}}%
\put(500,300){\makebox(0,0)[r]{1e-06}}%
\end{picture}%
\endgroup
 
\caption{Matter power spectra in the CDM-type hybrid models for high
values of the parameter $\lambda$. All the curves have been normalised
to COBE.  Note that one the power spectrum goes to zero for a critical
scale, which depends on the value of $\lambda$.}
\label{fig_cor3ps}
\end{figure}

To illustrate the power spectra corresponding to correlated hybrid
perturbations, we have plotted on
\FIGS{\ref{fig_cor1ps}-\ref{fig_cor3ps}}, the power spectra (taking
the gravitational potential perturbation as reference variable) in the
case of CDM-type correlated hybrid perturbations, for various values
of $\lambda$.
 
Some interesting behaviour occurs for large values of $\lambda$. When
$\lambda \geq \alpha$, it is easy to see using \EQ{\ref{theta_lambda}}
that the CDM and the baryon density contrasts have initially opposite
signs. The CDM density constrast evolves [see
\EQNS{\ref{dc_cons},\ref{dc_euler}}] according to the equation
\begin{equation}
\ddot \Dc + \h \dot\Dc = -k^2 \Phi,
\end{equation}
which can be solved to give
\begin{equation}
\Dc = \Dc^{\rm init} -k^2 
\int\left[\frac{1}{a}\int a\Phi \ddd \eta\right]\ddd \eta',
\end{equation}
where $\Dc^{\rm init}$ is the initial value for the CDM density
contrast, so that it will evolve towards values with sign opposite to
that of $\Phi$, that is of same sign as $\Da$. Moreover shortest
wavelengths will evolve rapidly enough and change sign whereas longest
wavelengths will not, so that the CDM power spectrum should exhibit a
sign change. For a given mode, this sign change occurs all the more
rapidly as the initial CDM density contrast is small as compared to
$\hat\Phi$, that is when $\lambda$ is small. Plotting the matter power
spectrum, it is therefore natural to expect that the wavenumber at
which it is zero is all the more small as $\lambda$ is small. This
is what we can check on \FIG{\ref{fig_cor3ps}}.

\subsection{CMB polarization}

\begin{figure}
\centering
% GNUPLOT: LaTeX picture with Postscript
\begingroup%
  \makeatletter%
  \newcommand{\GNUPLOTspecial}{%
    \@sanitize\catcode`\%=14\relax\special}%
  \setlength{\unitlength}{0.12bp}%
\begin{picture}(3600,2160)(0,0)%
\special{psfile=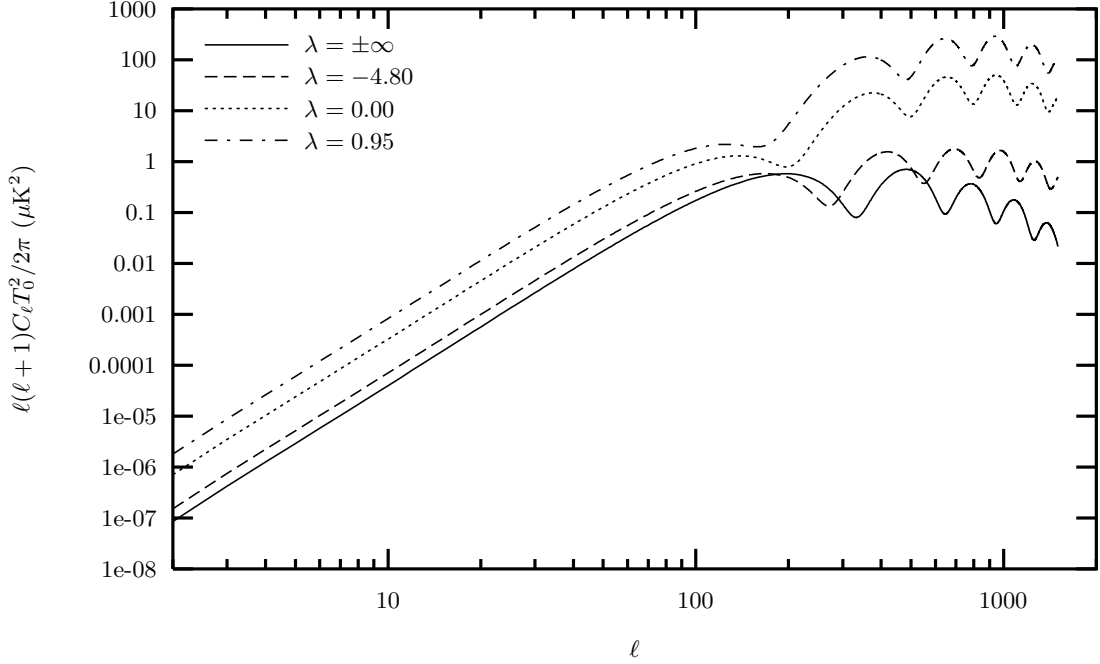 llx=0 lly=0 urx=600 ury=420 rwi=7200}
\put(963,1647){\makebox(0,0)[l]{$\lambda = 0.95$}}%
\put(963,1747){\makebox(0,0)[l]{$\lambda = 0.00$}}%
\put(963,1847){\makebox(0,0)[l]{$\lambda = -4.80$}}%
\put(963,1947){\makebox(0,0)[l]{$\lambda = \pm\infty$}}%
\put(2000,50){\makebox(0,0){$\ell$}}%
\put(100,1180){%
\special{ps: gsave currentpoint currentpoint translate
270 rotate neg exch neg exch translate}%
\makebox(0,0)[b]{\shortstack{$\ell(\ell+1) C_\ell T_0^2 / 2 \pi$ ($\mu$K$^2$)}}%
\special{ps: currentpoint grestore moveto}%
}%
\put(3159,200){\makebox(0,0){1000}}%
\put(2192,200){\makebox(0,0){100}}%
\put(1226,200){\makebox(0,0){10}}%
\put(500,2060){\makebox(0,0)[r]{1000}}%
\put(500,1900){\makebox(0,0)[r]{100}}%
\put(500,1740){\makebox(0,0)[r]{10}}%
\put(500,1580){\makebox(0,0)[r]{1}}%
\put(500,1420){\makebox(0,0)[r]{0.1}}%
\put(500,1260){\makebox(0,0)[r]{0.01}}%
\put(500,1100){\makebox(0,0)[r]{0.001}}%
\put(500,940){\makebox(0,0)[r]{0.0001}}%
\put(500,780){\makebox(0,0)[r]{1e-05}}%
\put(500,620){\makebox(0,0)[r]{1e-06}}%
\put(500,460){\makebox(0,0)[r]{1e-07}}%
\put(500,300){\makebox(0,0)[r]{1e-08}}%
\end{picture}%
\endgroup
 
\caption{CMB polarisation anisotropies in the CDM-type hybrid models
for various values of the parameter $\lambda$. We have represented the
spectrum for the same values of $\lambda$ as in
\FIG{\ref{fig_cor1dt}} (but we have omitted two of them for
clarity). The amplitude of the spectrum decreases and the peaks shift
to the right as one goes from the adiabatic case ($\lambda = 0$,
dotted line) to the pure isocurvature case ($\lambda = \pm\infty$,
solid line), as the spectrum closely follows that of the Doppler
contribution to the temperature anisotropy spectrum (see
\FIGS{\ref{fig_adi} and \ref{fig_iso}}).}
\label{fig_pol}
\end{figure}

\begin{figure}
\centering
% GNUPLOT: LaTeX picture with Postscript
\begingroup%
  \makeatletter%
  \newcommand{\GNUPLOTspecial}{%
    \@sanitize\catcode`\%=14\relax\special}%
  \setlength{\unitlength}{0.12bp}%
\begin{picture}(3600,2160)(0,0)%
\special{psfile=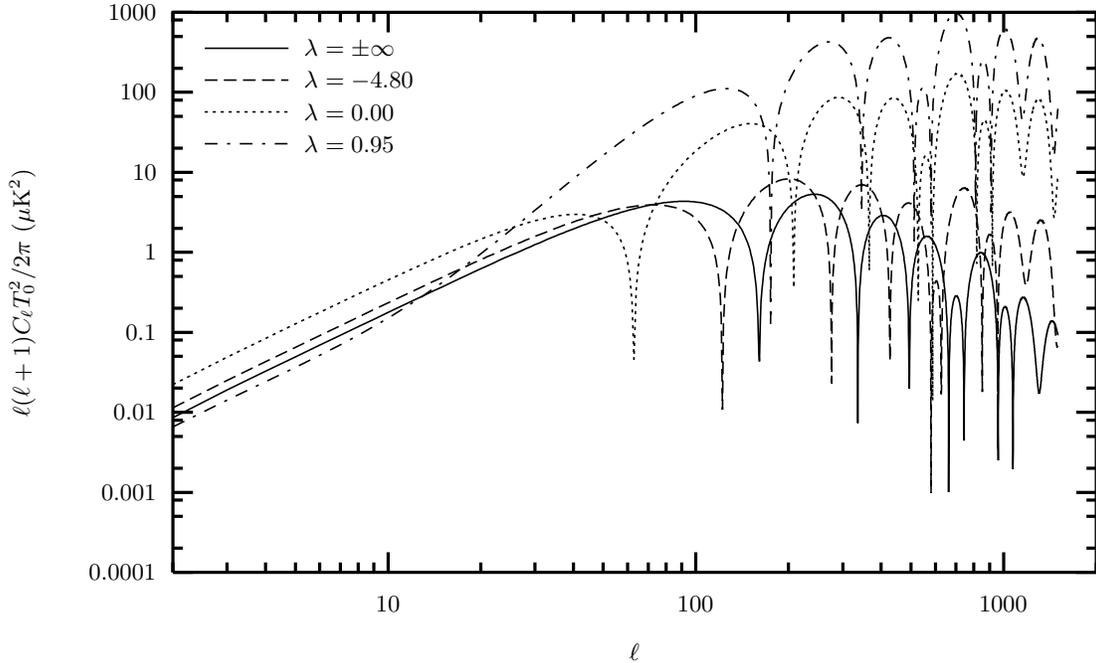 llx=0 lly=0 urx=600 ury=420 rwi=7200}
\put(963,1647){\makebox(0,0)[l]{$\lambda = 0.95$}}%
\put(963,1747){\makebox(0,0)[l]{$\lambda = 0.00$}}%
\put(963,1847){\makebox(0,0)[l]{$\lambda = -4.80$}}%
\put(963,1947){\makebox(0,0)[l]{$\lambda = \pm\infty$}}%
\put(2000,50){\makebox(0,0){$\ell$}}%
\put(100,1180){%
\special{ps: gsave currentpoint currentpoint translate
270 rotate neg exch neg exch translate}%
\makebox(0,0)[b]{\shortstack{$\ell(\ell+1) C_\ell T_0^2 / 2 \pi$ ($\mu$K$^2$)}}%
\special{ps: currentpoint grestore moveto}%
}%
\put(3159,200){\makebox(0,0){1000}}%
\put(2192,200){\makebox(0,0){100}}%
\put(1226,200){\makebox(0,0){10}}%
\put(500,2060){\makebox(0,0)[r]{1000}}%
\put(500,1809){\makebox(0,0)[r]{100}}%
\put(500,1557){\makebox(0,0)[r]{10}}%
\put(500,1306){\makebox(0,0)[r]{1}}%
\put(500,1054){\makebox(0,0)[r]{0.1}}%
\put(500,803){\makebox(0,0)[r]{0.01}}%
\put(500,551){\makebox(0,0)[r]{0.001}}%
\put(500,300){\makebox(0,0)[r]{0.0001}}%
\end{picture}%
\endgroup
 
\caption{CMB temperature-polarisation cross-correlation spectrum in
the CDM-type hybrid models for various values of the parameter
$\lambda$. As in\FIG{\ref{fig_pol}}, we have represented the spectrum
for the same values of $\lambda$ as in \FIG{\ref{fig_cor1dt}} and have
omitted two of them for clarity. The amplitude of the spectrum
decreases and the peaks shift to the right as one goes from the
adiabatic case ($\lambda = 0$, dotted line) to the pure isocurvature
case ($\lambda = \pm\infty$, solid line).}
\label{fig_polcross}
\end{figure}

Although CMB polarization has not yet been measured and is expected to
be difficult to measure, it can provide a lot of additional
information concerning the cosmological perturbations and the
cosmological parameters~\cite{zs97}. This is why we will consider
briefly the consequence of correlated hybrid perturbations on the CMB
polarization. The polarization can be decomposed~\cite{zs97} into
$E$-mode polarization and $B$-mode polarization, and we shall define,
in addition to the temperature angular power spectrum, the $E$-mode
angular power spectra,
\beq
C_\ell^{EE}\equiv \langle |a^E_{\ell m}|^2\rangle,
\eeq
and the correlation spectrum 
\beq
C_\ell^{ET}\equiv\langle a_{\ell m}^{T*}a_{\ell m}^E\rangle,
\eeq
where the $a_{\ell m}^{T}$ correspond here to the same coefficients as
those defined in (\ref{multipole}), which gives the correlation
between the temperature and the $E$-mode polarization. Scalar-type
perturbations do not contribute to the $B$-mode polarization. As an
illustration, we have plotted the $E$-mode polarisation anisotropy
spectrum for various values of $\lambda$. Two features in these curves
are obvious. First, the amplitude of the polarisation varies, as well
as the position of the first peak in the spectrum. This can be
understood by looking at the Boltzmann equation for the photon
fluid. Before the last scattering surface, Thomson diffusion in
important, and the photons anisotropic stress approximatively obeys to
the equation~:
\begin{equation}
\sigma_\gamma \simeq - \frac{4}{15}\frac{k}{\dot \kappa}\Vbg ,
\end{equation}
where $\dot\kappa$ stands for the differential Thomson opacity, and
$\Vbg$ is the photon-baryon plasma velocity. Thus, the photons
anisotropic stress (which is proportional to the $E$-type
polarization, see \EQNS{62,63,77} of~\cite{hw97}) is proportional the
photon dipole, and the $C_\ell^{EE}$ are proportional to the Doppler
contribution of the CMB temperature anisotropy spectrum. Therefore,
the position and height of the peaks vary for both spectra in the same
way with $\lambda$. These results are represented in
\FIG{\ref{fig_pol}} and the cross-correlation spectrum between
temperature and polarisation can be found on \FIG{\ref{fig_polcross}}.

\section{Conclusion}
\label{sec_conc}

In the present work, our goal has been to analyze the effects of
correlated hybrid adiabatic and isocurvature perturbations on
observational quantities, in particular the CMB anisotropies. We have
isolated four ``elementary'' modes, corresponding to a deviation from
adiabaticity of one of the four standard species (photons, neutrinos,
baryons, CDM)~: each type of these elementary modes is characterized
by only two parameters ($S$ and $\Da$).  One could, of course,
generalize our analysis by relaxing adiabatic ratios among the three
remaining species, but at the price of requiring four parameters.

We have shown, in the case of these elementary modes, that correlation
leads to very specific effects on the CMB anisotropies and on large
scale structure which do not appear in the case of independent
mixtures.

In this paper, our purpose was not to confront directly observations
with this type of models, but rather to focus on some qualitative
interesting consequences of correlated hybrid perturbations.  This is
why we have considered only a subclass of models, which are extreme in
the sense that they are totally correlated and simple because they are
described by only two parameters. If one wishes to compare hybrid
models with observations, one should consider the {\it sum} of spectra
of the type we have obtained (as explained in \S\ref{sec_corr}), which
means that the adiabatic and isocurvature perturbations would then be
only partially correlated (this is the case in the specific model
of~\cite{l99}).

The present data are still too imprecise to be able to distinguish
this kind of correlated hybrid perturbations, but it may be
interesting to know how much precise data would constrain these modes.
In practice, it might turn out to be a difficult task to disentangle
the presence of such modes in the data, unless one assumes a specific
early universe model.

\acknowledgments

It is a pleasure to thank Nabila Aghanim, Francis Bernardeau, Martin
Bucher, Kavilan Moodley and David Spergel for interesting discussions.

\appendix

\section{Evolution of perturbations}
\label{sec_appendix} 

In this appendix, we derive the evolution of all quantities for long
wavelengths, \IE{} {\it scales outside the Hubble radius}~: $k\ll aH$.
The whole system of equations, in a notation slightly different from
the one adopted here, can be found for example in \cite{mb95}.

Let us first introduce the system of equations governing the evolution
of the matter perturbations, which related the density contrasts of
the four species to the scalar component of their velocities (denoted
by $V_A$). We first have four equations of conservation, for each of
the four species, that read (in Fourier space)~:
\begin{eqnarray}
\dot\Dn &=& \frac{4}{3}{k }\Vn \cr
\label{dc_cons}
\dot\Dc &=& {k }\Vc \cr
\dot\Dp &=& \frac{4}{3}{k }\Vbg \cr
\dot\Db &=& {k }\Vbg,
\end{eqnarray}
where a prime denotes a derivative with respect to the conformal time
and where $\Vbg$ is the velocity common to the baryon and photon
fluids, which are coupled until last scattering.  We then have three
Euler equations, two for the independent fluids of CDM and neutrinos,
and one for the baryon-photon fluid~:
\begin{eqnarray}
\dot\Vn &=& -{k }\left[{\Dn\over 4}+\Psi+\Phi - \sigma_\nu\right], \cr
\label{dc_euler}
\dot\Vc &=& -{\h\Vc } -{k }\Phi, \cr
\dot\Vbg &=& -{3 \Omb\over 4\Omp+3\Omb }\h{\Vbg} 
-{k }{4\Omp\over 4\Omp+3\Omb } 
\left[\frac{\Dp}{4} +\Psi\right] -k\Phi,
\label{euler}
\end{eqnarray}
where $\h$ is the comoving Hubble parameter defined by $\h\equiv
a'/a$.  In the last equation, the coefficients $\Omega_A$ are time
dependent since the ratio of the energy density of a given species
with respect to the critical energy density, will change with time.

To close the above system of equations, one needs the Einstein
equations, which expresse the metric perturbations in terms of the
matter perturbations.  Only two components of the Einstein equations
are useful, the other ones being redundant, and they are the Poisson
equation
\beq
-\left[{k^2\over \h^2}+\frac{9}{2}(1+w)\right]\Psi=\frac{3}{2}\sum_X
\Omega_X\left[\D_X-{3\h\over k}(1+w_X)\W_X\right],
\label{poisson}
\eeq
and the anisotropic stress equation,
\beq
{k^2\over \h^2}\left(\Psi-\Phi\right)=6\Omn\sigma_\nu, \label{anisotropy}
\eeq
where $\sigma_\nu$ represents the anisotropic stress due to the
neutrinos (which thus require a decription beyond the perfect fluid
approximation).  To get the evolution of $\sigma_\nu$, one must use a
higher moment of the Boltzmann equation (see \cite{mb95}),
\beq
\dot\sigma_\nu=-\frac{4}{15}k\Vn, \label{evol_anisotropy}
\eeq
where other terms on the right hand side have been neglected.

In order to solve the above equations for long wavelengths, it is
convenient to consider an expansion of all quantities in terms of the
small parameter $k\eta$, so that one will have
\beq
X=X^{(0)}+X^{(1)} k\eta +X^{(2)} (k\eta)^2 +\dots
\eeq
The Euler equations (\ref{euler}) then enable us to express the first
order velocity term as a function of the zeroth order density and
gravitational potentials,
\begin{eqnarray}
\Vno&=&-\frac{1}{4}\Dno-\Psio-\Phio,\cr
\Vco &=& -{\Phio\over 1+\h \eta},\cr
\Vbgo&=&-{\Omp\Dpo+4\Omp\Psio+(4\Omp+3\Omb)\Phio\over 4\Omp+3(1+\h\eta)\Omb}.
\end{eqnarray}
The zeroth order components of the velocities are as usual set to
zero, otherwise one would get a divergence in the right hand side of
(\ref{poisson}) (unless there is a special cancellation of the type
mentioned in \cite{bmt99}).

Using (\ref{anisotropy}) and (\ref{evol_anisotropy}) at lowest order,
one gets
\beq
\Psio-\Phio=-\frac{4}{5}\Omn\h^2\eta^2\Vno,
\eeq
and susbtituting the above expression for $\Vno$, one finds the
following relation between $\Phio$ and $\Psio$~:
\beq
\left(1+\frac{4}{5}\Omn\h^2\eta^2\right)\Phio=
\left(1-\frac{4}{5}\Omn\h^2\eta^2\right)\Psio
-\frac{1}{5}\Omn\h^2\eta^2 \Dno.
\label{phipsi}
\eeq
Substituting in Poisson's equation (\ref{poisson}), at lowest order,
the expressions obtained above for the velocities, and using
(\ref{phipsi}), one finally gets the following cumbersome equation,
relating $\Phio$ to the four species densities,
\begin{eqnarray}
&-&\left\{(3+\Omp+\Omn)\left(1+\frac{4}{5}\Omn y^2\right) 
+{(4\Omp+3\Omb)y\over 4\Omp+3\Omb(1+y)}\left[8\Omp+3\Omb 
\left(1-\frac{4}{5}\Omn y^2\right)\right] \right. \cr
&& \left. +{3y\over 1+y}\Omc
\left(1-\frac{4}{5}\Omn y^2\right)+8y\Omn\right\}
{\Phio\over 1-\frac{4}{5}\Omn y^2} \cr &=&
\Omb\Dbo+\Omc\Dco+{4\Omp(1+y)+3\Omb(1+2y)\over
4\Omp+3\Omb(1+y)}\Omp\Dpo \cr &+& \left[\frac{4}{5}\Omp
y^3{4\Omp+3\Omb\over 4\Omp+3\Omb(1+y)} +1+y+(3+\Omp-3\Omn){y^2\over
5}\right] {\Omn\Dno\over 1-\frac{4}{5}\Omn y^2},
\end{eqnarray}
where $y\equiv \h\eta$.  While this expression yields the evolution of
the gravitational potential perturbation during the whole evolution of
the universe from the deep radiation era till the last scattering, it
will be sufficient for our present purpose to retain from this
equation only its asymptotic forms in the radiation era and the matter
era.  In the radiation era , $y=1$ and $\Omc, \Omb \ll \Omp, \Omn$, so
that the above expression simplifies to give
\beq
\Phio=-\frac{1}{4}\left(3+\frac{4}{5}\Omn\right)^{-1}\left[
2\left(1-\frac{4}{5}\Omn\right)\Omp\Dpo
+\frac{2}{5}\left(9-4\Omn\right) \Omn\Dno\right].
\eeq
In the matter era, $y=2$ and $\Omp, \Omn \ll\Omc, \Omb$, so that one
finds
\beq
\Phio=-\frac{1}{5}\left(\Omb\Dbo+\Omc\Dco\right).
\eeq

\figure

\end{document}